\documentclass[amsfonts,amsmath,prd,preprint,nofootinbib]{revtex4}
\newcommand{\beq}{\begin{equation}}
\newcommand{\eeq}{\end{equation}}

\newcommand{\bsp}{\begin{split}}

\usepackage{epsfig,bbm,cancel,ulem}
\usepackage[breaklinks=true]{hyperref}
\usepackage{xcolor}
\usepackage{wasysym}
\usepackage[mathscr]{eucal}
\usepackage{multirow}
\usepackage{amsmath}
\usepackage{enumitem}

\begin{document}

\title{Bound orbits around charged black strings}

\author{A S. Habibina}
%\email{a.sayyidina@sci.ui.ac.id}
%\author{B.~N. Jayawiguna}
%\email{byon.nugraha@ui.ac.id}
\author{H.~S.~Ramadhan}
%\email{hramad@sci.ui.ac.id}
\affiliation{Departemen Fisika, FMIPA, Universitas Indonesia, Depok, 16424, Indonesia. }
\def\changenote#1{\footnote{\bf #1}}

\begin{abstract}
We study the geodesics of $5d$ Reissner-Nordstrom and nonsingular black strings, and establish a rational bound orbit taxonomy for both massive as well as null test particles. For the timelike case, test particles with high energy (that would have made them plunge into or scatter off a black hole) could still form bound orbits around the black strings. We calculate the accumulated angles of the corresponding radial periods and show that they are higher than their $4d$ counterparts. For the null case, we found the existence of stable null orbits outside their respective horizons, which do not exist in the four dimensions except at their extremal limit.
\end{abstract}

\maketitle
\thispagestyle{empty}
%\section{Introduction}
\setcounter{page}{1}

\section{Introduction}
\label{intro}

Black holes in higher dimensions have long been an interesting as well as intriguing subject. Mathematically, the additions of extra dimensions pose a considerable challenge to obtaining exact solutions. Astrophysically, their signature might shed light on the unification of fundamental forces. Although the possibility of extra dimension has been considered by Kaluza and Klein as early as the 1920s~\cite{Kaluza:1921tu}, it was Tangherlini who generalized the Reissner-Nordstrom (RN) black hole to higher dimensions~\cite{Tangherlini:1963bw}. Myers and Perry then followed in obtaining solutions for higher-dimensional Kerr-Newman~\cite{Myers:1986un}. The spherical event horizon is not the only possible higher-dimensional BH solution. A trivial extension of a Schwarzschild black hole to a higher dimension would be a charged black string whose metric does not depend on its compact extra dimension. Such solution has been shown to be stable by Gregory and Laflamme~\cite{Gregory:1987nb}. More nontrivial black strings are, of course, abundant in the literature~\cite{Gibbons:1987ps, Gregory:1995qh, Emparan:2001wn}. 

Even in $4d$ due to its very definition, a black hole cannot be observed directly. Thus, so far the probe for their existence relies on: (i) the gravitational wave~\cite{LIGOScientific:2016aoc} and (ii) orbital dynamics around it (e.g., perihelion precession~\cite{GRAVITY:2020gka}, gravitational lensing and shadow~\cite{EventHorizonTelescope:2019dse, Akiyama2022}, or photon orbits~\cite{GRAVITY:2018}). It was Hagihara in 1930 that first obtained the exact solution tof the Schwarzschild geodesic~\cite{Hagihara(1930)} (see also Hackmann~\cite{Hackmann:2010tqa}). The analytic solution is in the form of the Weierstrass elliptic function. It was later extended to the case of (static and rotating) black strings by Grunau~\cite{Grunau:2013oca}. The behavior of massive neutral and charged particles around weekly magnetized Schwarzschild string was studied~\cite{Rezvanjou:2017hox}. 

In a series of interesting papers, Levin and collaborators propose a taxonomy based on isomorphism between periodic orbits around (or of pairs of) $4d$ black holes and rational numbers~\cite{Levin:2008mq, Levin:2008ci, Perez-Giz:2008ajn, Healy:2009zm}. They found that there exists a set of three integers $(z, w, v)$ that can be combined to form a rational number
\begin{equation}
\label{qnumber}
q\equiv w+{v\over z},
\end{equation}
that completely parametrizes every (timelike or null) bound orbit. Here $w$ defines the number of whirls, $z$ expresses the number of leaves, and $v$ denotes the order the leaves are traced out. Within this formalism, precession of Mercury's perihelion can be perceived as periodic orbit with very large $z$, around $z\sim1.296\times 10^7$~\cite{Levin:2008mq}. This rational orbit formalism has been applied to Schwarzchild, RN~\cite{Misra:2010pu}, and Kerr solutions~\cite{Levin:2008mq}, as well to the black hole binary system.

%\textcolor{red}{To the best of our knowledge, this rational orbit formalism has not been applied to the case of black strings. If orbital dynamics can probe the existence of black hole, }

%It has long been realized that RN is not the only charged solution in the static limit. Bardeen showed that a black hole can be made regular (devoid of any singularity inside its horizon) by charging it with electric/magnetic charge~\cite{bardeen}. Unfortunately, the Bardeen black hole also lacks stable photon orbits outside its horizon. It should be noted that, to the best of our knowledge, there has yet been any discussion on regular black strings in the literature. 
\textcolor{red}{}%\textcolor{red}{Connecting this problem with previously mentioned studies, we find that a uniform black string might be the possible scenario to crack this issue. In her work, Grunau has shown that bound orbits of light are possible both in the static (Schwarzchild) and the rotating (Kerr) black string spacetime. By investigating her method, we believe that extending charged $4d$ black hole into $5d$ black string might give more enlightenment toward our pursuit of discovering null bound orbits in charged spacetime.}

%\textcolor{red}{The $4d$ geodesic discussed above teaches a lesson. If we study the orbital dynamics around some black hole and obtain different values for observable}

Even though its unchraged geodesic has extensively been studied~\cite{Grunau:2013oca}, to the best of our knowledge no one yet elaborates the rational orbit formalism on the charged black string. It is therefore the purpose of this paper to accomplish this task. Since the orbital motion is confined to $4d$, the hope is that the effect of one extra spatial dimension will give distinct observational signatures that can distinguish black hole from black string. The ``anomaly" orbital motion around a black object can thus serve as an astrophysical probe for the existence of extra dimension. To do so, we specifically study two types of charged strings: the charged Gregory-Laflamme~\cite{Gregory:1987nb} and the nonsingular black strings. The latter is the $5d$ extension of the well-known Bardeen black hole~\cite{bardeen}. These two are simple toy models but at the same time are rich enough to prove our point. The organization is as follows. We examine the general black string geodesic and effective potential in Section~\ref{bsg}. Section~\ref{orbit} is devoted to the all possible types of orbits and the formalism of rational orbit taxonomy. The orbital zoo of RN and Bardeen black strings are presented in Sections~\ref{rn} and ~\ref{b}, respectively. Finally, we conclude the result of this study in \ref{con}.

\section{Black String Geodesics}
\label{bsg}

A trivial $5d$ extension of a static and spherically-symmetric black hole can always be written as~\cite{Gregory:1987nb}
\begin{equation}
\label{bs}
ds^2 = -f(r) dt^2 +f(r)^{-1} dr^2 + r^2 d\Omega^2 + d\omega^2,
\end{equation}
where $\omega$ is an extra compact spatial dimension and $f(r)$ is any metric solution that solves the $5d$ Einstein equations. Since the string preserves the $4d$ spherical symmetry, under the condition of equatorial plane ($\theta=\pi/2$) the invariance condition $g_{\mu\nu}\dot{x}^{\mu}\dot{x}^{\nu}=-\epsilon$ gives the constraint equation as
\begin{equation}
\label{el}
-\epsilon = -f(r) \dot{t}^2 +f(r)^{-1} \dot{r}^2 + r^2 \dot{\phi}^2 +\dot{\omega}^2,
\end{equation}
where $\epsilon=\{1,0\}$ corresponds to timelike and null geodesics, respectively. Using Euler-Lagrange equation, we obtain the equation of motion for each coordinate:
\begin{eqnarray}
\label{tdot}
\dot{t} &=& \frac{dt}{d\tau} = \frac{\mathbb{E}}{f(r)}, \nonumber\\
%\label{pdot}
\dot{\phi} &=& \frac{d\phi}{d\tau} = \frac{\mathbb{L}}{r^2}, \nonumber\\
%\label{wdot}
\dot{\omega} &=& \frac{d\omega}{d\tau} = \mathbb{J}.
\end{eqnarray}
Inserting \eqref{tdot} into \eqref{el} and rescaling $\mathbb{L}^2 \rightarrow 1/L$, $\mathbb{J}^2\rightarrow J$, we have the geodesic equation in the form of
\begin{equation}
\label{geo}
\dot{r}^2 = \mathbb{E}^2 - V_{eff},
\end{equation}
where the effective potential $V_{eff}$ can be written as
\begin{eqnarray}
\label{Veff}
V_{eff}(r)&=& f(r)\left(\epsilon +J + \frac{1}{L r^2} \right).
\end{eqnarray}
%We perform the same rescale $r$-equation and 

\section{Black String Orbits}
\label{orbit}

\subsection{Types of Orbit}

\begin{figure}[!ht]
	\centering
	\begin{tabular}{cc}
		\includegraphics[height=2.6cm]{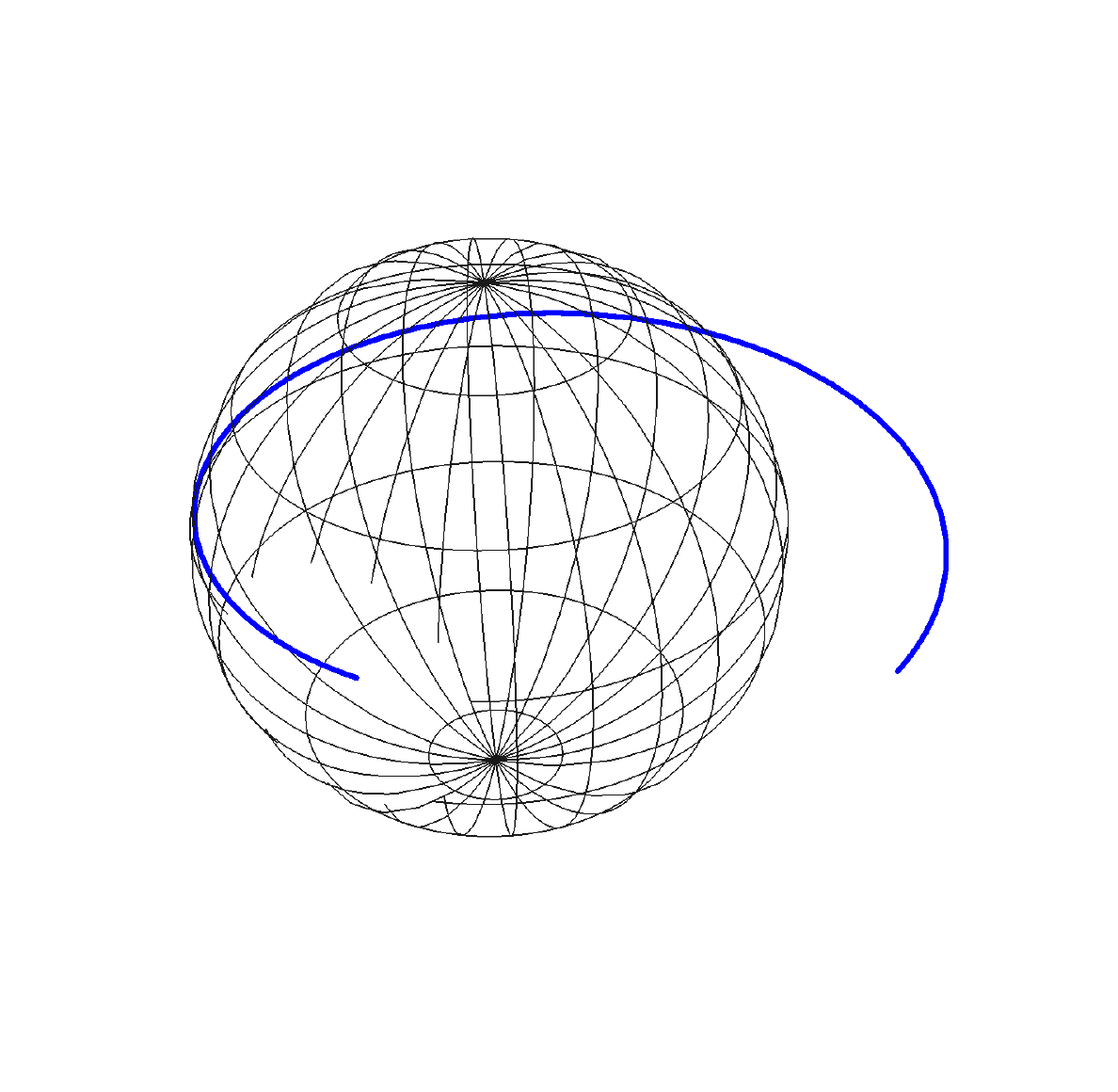}&
		\includegraphics[height=3.1cm]{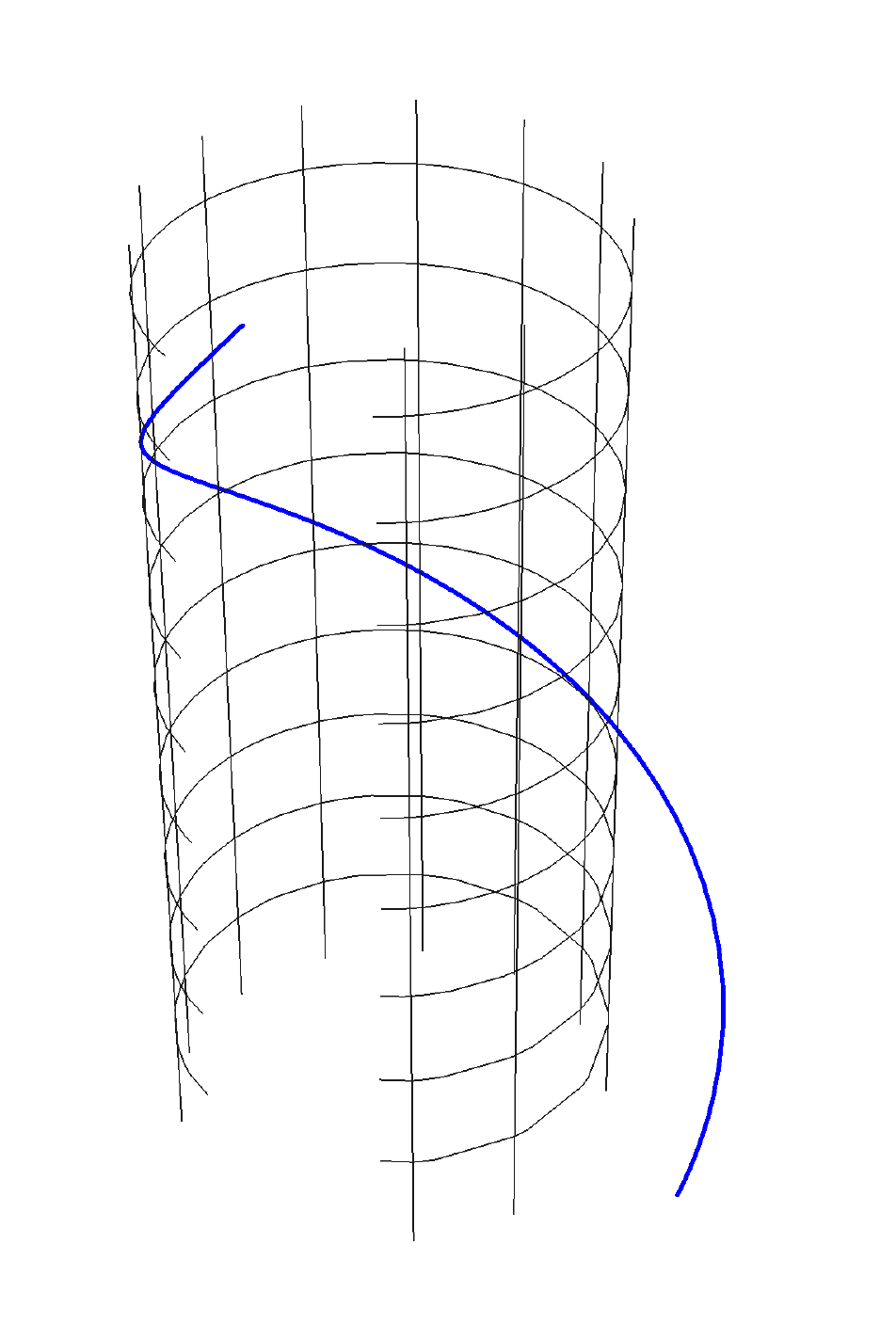} \\
		\includegraphics[height=3.6cm]{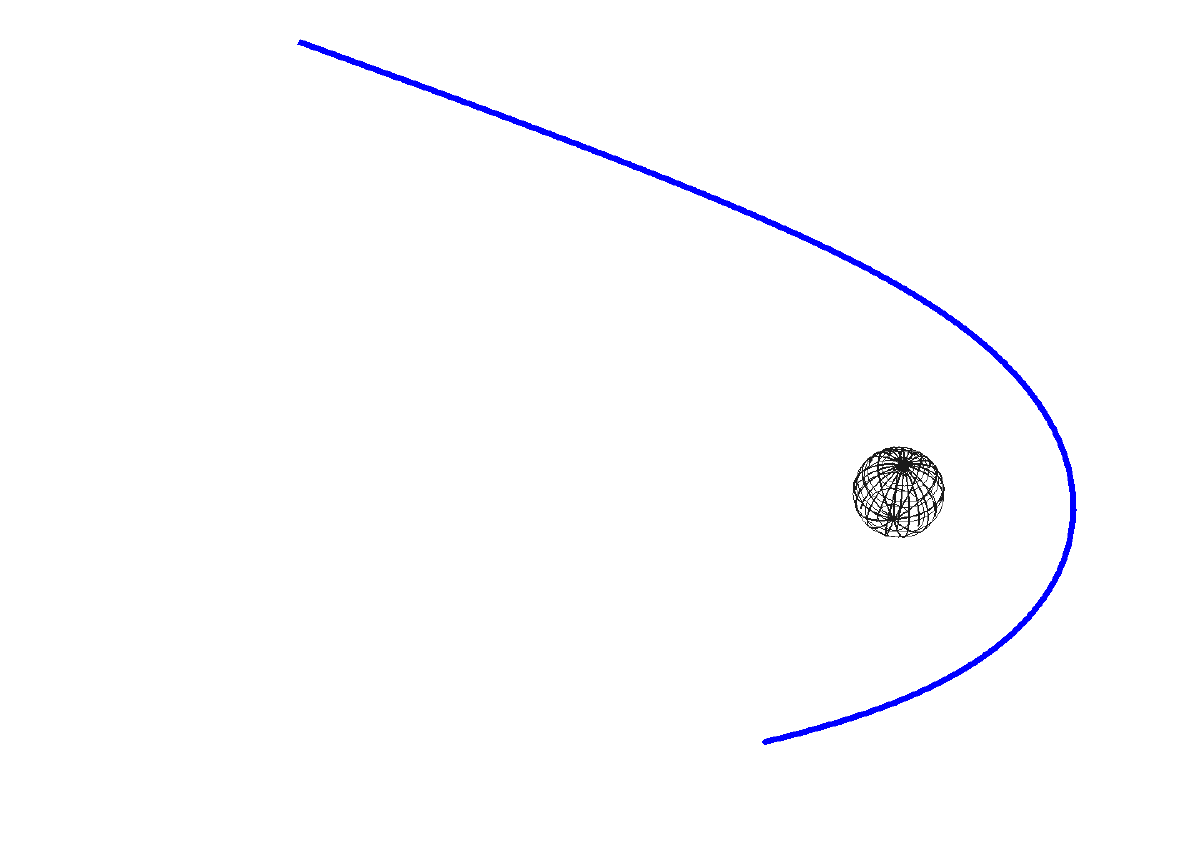}&	\includegraphics[height=4.1cm]{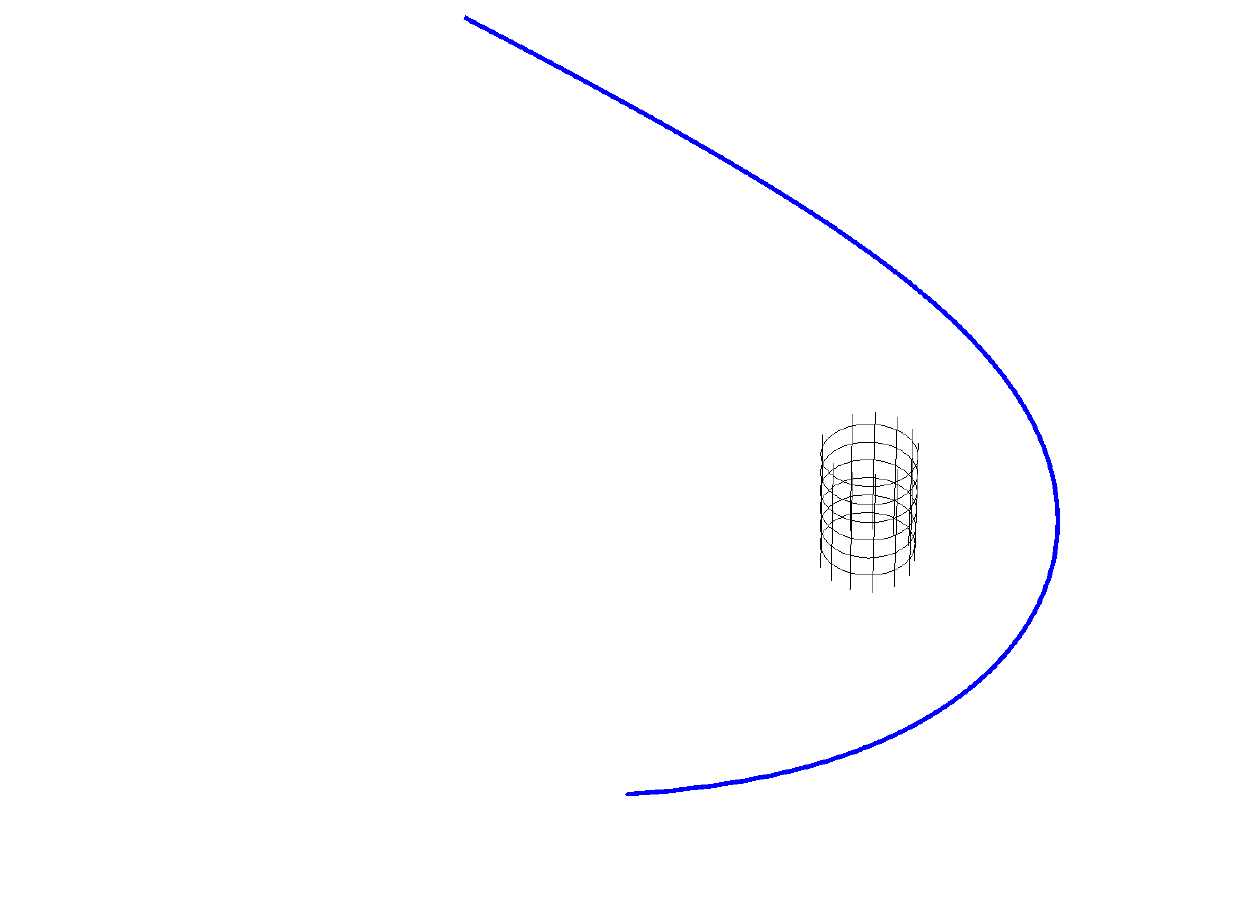} \\
		\includegraphics[height=5.6cm]{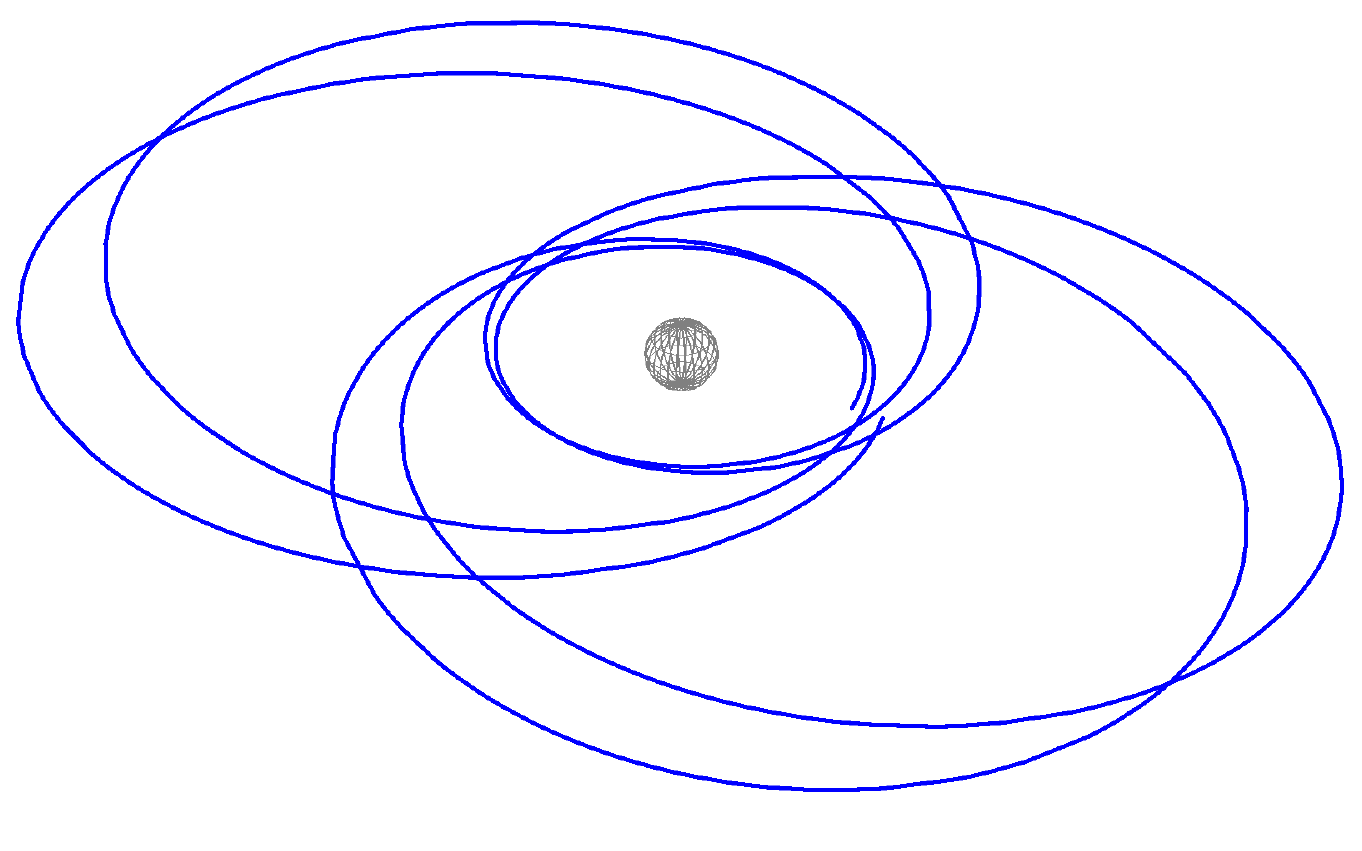}&	\includegraphics[height=6.1cm]{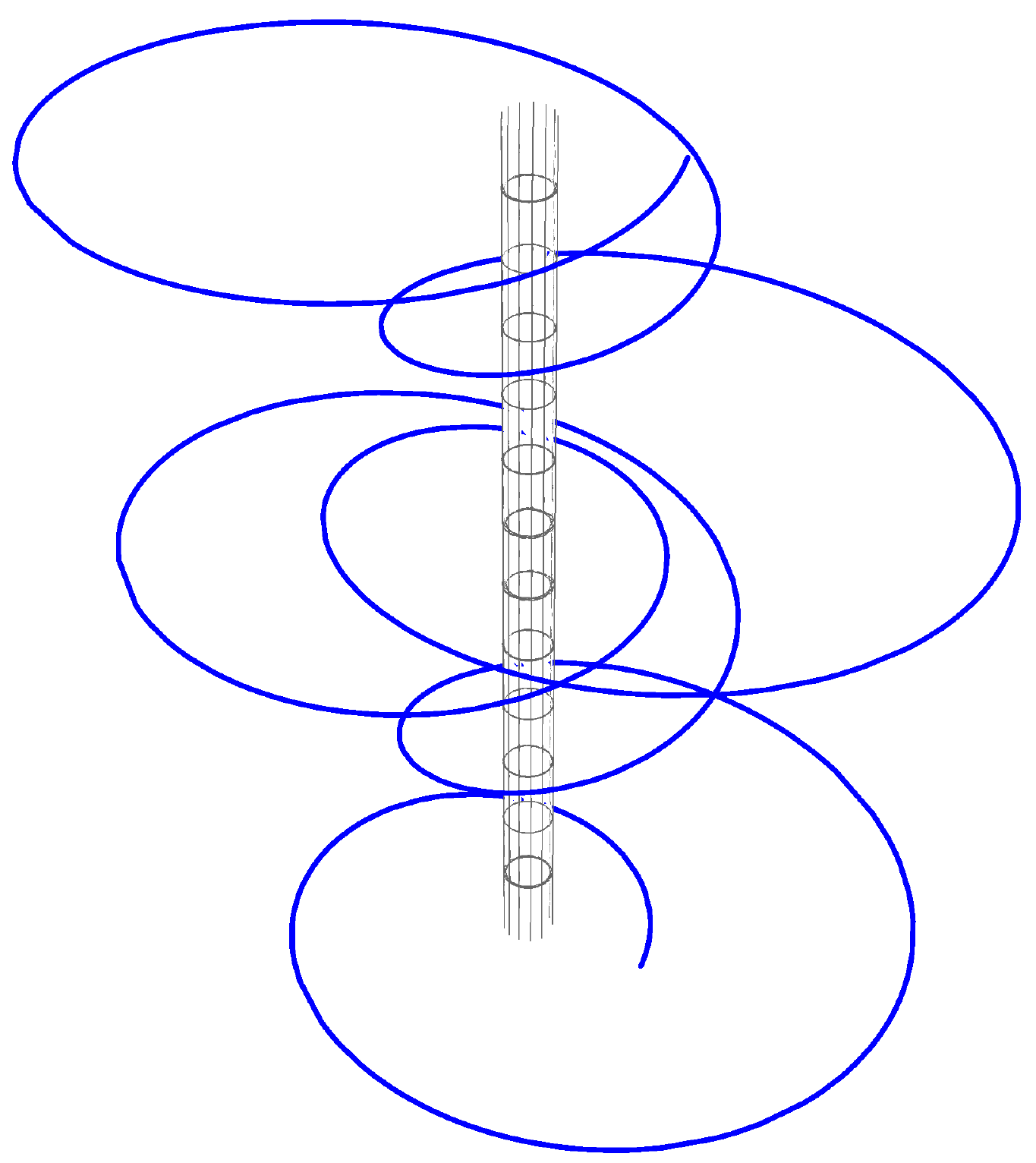}
	\end{tabular}			
	\caption{Typicals of terminating orbit, escape orbit, and bound orbit around black hole (left column) and black ring (right columnn).}
	\label{3d}
\end{figure}

Particle trajectories in black string can be classified in terms of orbital types and regions. Denoting $r_{EH}$ as the radius of outer event horizon, the three possible orbit types and their definitions are listed as below, in which their visualizations are shown in Fig.~\ref{3d}. Here we use similar convention as in Grunau~\cite{Grunau:2013oca}. 
\begin{enumerate}[label=\arabic*., itemsep=0pt, topsep=0pt]
	\item {\it Terminating orbit} (TO): particles arrive at a periapsis $r_p$ and falls into singularity.
	\item {\it Escape orbit} (EO): particles come from $\infty$, approach a periapsis $r_p$, and return to $\infty$.
	\item {\it Bound orbit} (BO): particles oscillate between its periapsis and apoapsis under the condition of $r_{EH} < r_p < r_a < \infty$.
%\label{region}
\end{enumerate}
\begin{figure}[!ht]
	\centering
	\begin{tabular}{c}
		\includegraphics[height=6.5cm]{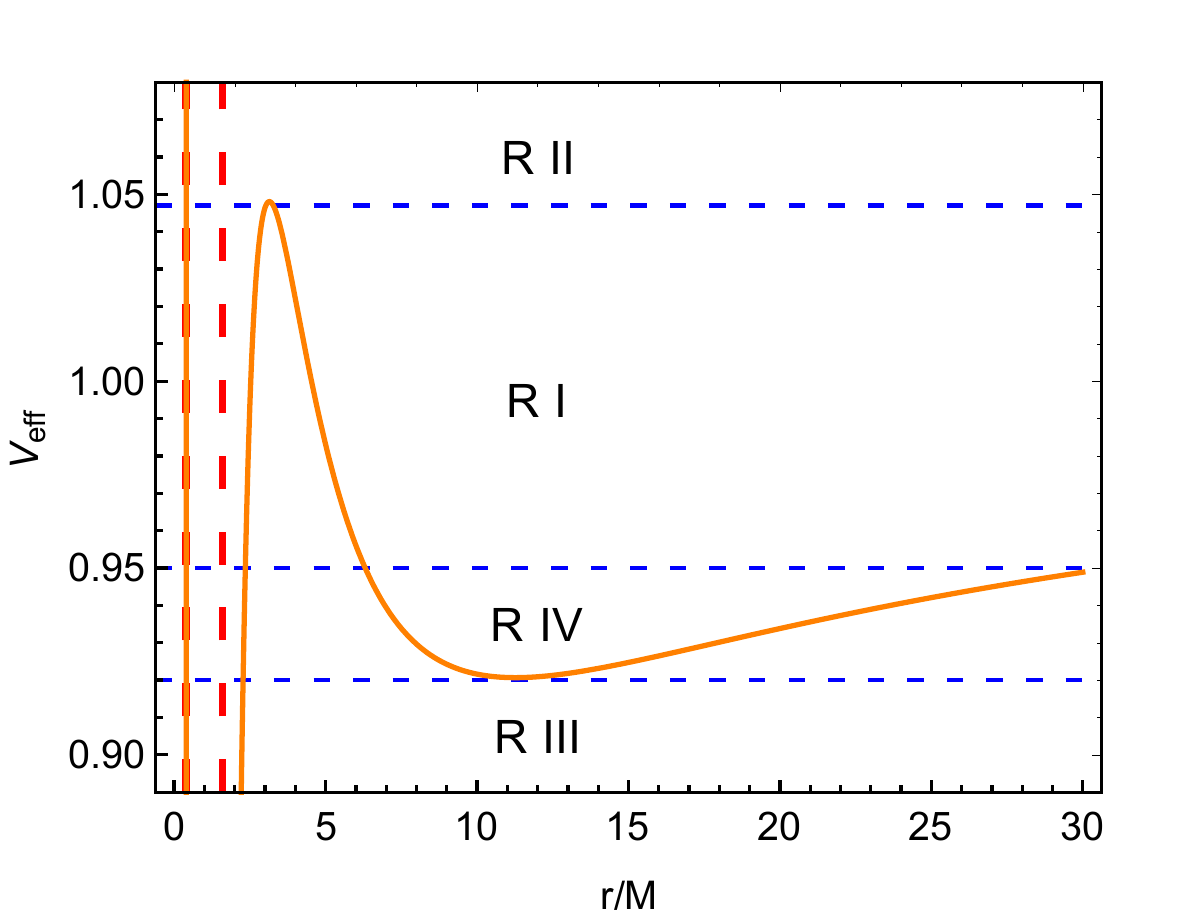}
	\end{tabular}
	\caption{All possible regions (R I to R IV) according to various energy level. The blue line refers to energy level $\mathbb{E}^2$.}
	\label{region}
	%\label{visual}
\end{figure}

\begin{table}[!hb]
	\setlength{\tabcolsep}{1.5em}
	\begin{tabular}{||c | c | c ||} 
		\hline
		Region & Positive Real Zeros & Orbit Types \\ [0.5ex] 
		\hline\hline
		I & 2 & TO, EO \\ 
		\hline
		II & 0 & EO  \\
		\hline
		III & 1 & TO \\
		\hline
		IV & 3 & TO, BO \\
		\hline
	\end{tabular}
	\caption{The classification of regions based on its effective $V_{eff}$ potential and particle energy level $E$ along with their corresponding possible orbit types.}
	\label{reg}
\end{table}
Based on the effective potential, the regions can be classified using the Hackman formalism~\cite{Hackmann:2010tqa}. Solving the Eq.~\eqref{geo} results in various scenarios based on the numbers of its positive real solutions, which are shown in Table~\ref{reg}. To help understanding the regions better, the dependence of the orbits on the energy levels is shown in Fig. \ref{region}. 

A typical $V_{eff}$ we study here generically have three turning points, as we shall see later, only two of which are outside the horizon. The local maxima represent the {\it unstable circular orbit} (UCO)\footnote{For timelike case, this local maxima is called the {\it marginally bound orbit} (MBO)~\cite{Ono:2015jqa}.}, while the local minima identify the {\it stable circular orbit} (SCO). Both circular orbits (CO) satisfy
\begin{equation}
\dot{r}^2=0,\ \ \ \partial_r V_{eff}=0,%\rightarrow V_{eff}=E^2,\nonumber\\
\end{equation}
which translates into
\begin{eqnarray}
\label{ucosco}
f(r_{CO})\left(J+\epsilon+{1\over L r_{CO}^2}\right)&=&\mathbb{E}^2,\nonumber\\
f'(r_{CO})\left(J+\epsilon+{1\over L r_{CO}^2}\right)-{2 f(r_{CO})\over L r_{CO}^3}&=&0.\nonumber\\
\end{eqnarray}
Both CO depend on the angular momenta and black hole's charge. For a given black hole there exists a critical circular radius beyond which it no longer supports bound orbits. For timelike case this critical value is called the {\it innermost stable circular orbit} (ISCO), and this is achieved when UCO and SCO radii merge. Mathematically, this is the inflection point of the effective potential,
\begin{equation}
\label{isco}
V_{eff}=\mathbb{E}^2,\ \ \ \partial_r V_{eff}=0,\ \ \ \partial^2_r V_{eff}=0.
\end{equation}
This adds the following condition,
\begin{equation}
r \left[r f''(r) \left(L r^2 (J+\epsilon )+1\right)-4 f'(r)\right]+6 f(r)=0,
\end{equation}
into~\eqref{ucosco}.

\subsection{Overview of Rational Orbits Taxonomy}
\label{taxo}

Here we briefly review the scheme of Levin and Perez-Giz in assigning each orbital solution to a distinct rational number~\cite{Levin:2008mq}.  Any (positive) rational number can always be cast as
\begin{equation}
q=s+{m\over n},
\end{equation}
where $s\geq0$ is an integer and $1\leq m\leq n-1$ are (relatively) prime numbers. This taxonomy draws correspondence between $q$ and the topological features of orbit. The claim is that each periodic orbit is completely characterized by three integers $(z, w, v)$ that can be expressed as rational number, Eq.~\ref{qnumber}. Each of them corresponds to a specific topological trait of a periodic orbit. The first and most tangible number is $z$, which is the number of leaves (or "zooms") of the orbit. The second integer, $w$, is the number of whirls which the particle performs in its path from apastron to periastron to the next apastron. Note that every particle performs at least a full $2\pi$ trip. The number of whirls is the additional integer number of full $2\pi$ accomplished beyond this. The last number $v$, the vertex number, differentiate between orbits that have equal $z$ and $w$ but are geometrically distinct. Particle can skip leaves in its motion from apastron to apastron. One important thing to address is the degeneracy that arises when the quotient $v=z$ is a reducible fraction. This is solved by requiring $v$ and $z$ to be relatively prime. The bounds on v then can be written as
\begin{eqnarray}
&1 \leq v \leq z-1	\,\,\,\,	&\text{if} \,\,\,\,\,\,\,\, z > 1 \,\, \text{and} \,\, z,v \,\, \text{are relatively prime} \nonumber \\
&v=0	\,\,\,\,	&\text{if} \,\,\,\,\,\,\,\, z=1.
\end{eqnarray}

The proof of such claim is based on the fact that for a periodic orbit, the accumulated angle between one apastron to another apastron can be expressed as
\begin{equation}
\label{acc_angle}
\Delta\varphi_r=2\pi\left(1+q\right)={\Delta\varphi\over z},
\end{equation}
where $\Delta\varphi\equiv z\Delta\varphi_r$ is the total accumulated angle in a complete orbit. We may also define the radial period as the (affine) time taken by particle to return to the same radius upon starting at the apsis $r_a$ (where $\mathbb{E}^2 = V_{eff}$)~\cite{Levin:2009sk, Babar:2017gsg}. If within this period, the evolution of $\varphi$ is an integer multiple of $2\pi$, we have a periodic orbit. Therefore the accumulated radial angle can be written as
\begin{equation}
\label{radangle}
\Delta\varphi_r= 2\int_{r_p}^{r_a}\frac{\dot{\phi}}{\dot{r}}dr = 2\int_{r_p}^{r_a} \frac{dr}{Lr^2\sqrt{\mathbb{E}^2-f(r)\left(\epsilon +J + \frac{1}{L r^2} \right)}},
\end{equation}
where $r_p$ is the periastron radius. The rational number in Eq.\eqref{acc_angle} can be related to the orbital angular frequency by the following arguments. Every eccentric equatorial orbit has two types of frequencies: the radial ($\omega_r$) and angular ($\omega_{\varphi}$) frequencies. They are given by
\begin{eqnarray}
\omega_r&=&{2\pi\over T_r},\nonumber\\
\omega_{\varphi}&=&{1\over T_r}\int_{0}^{T_r}\frac{d\varphi}{dt}dt={\Delta\varphi_r\over T_r},
\end{eqnarray}
where $T_r$ is the period of one radial cycle. Any periodic orbit must then satisfy
\begin{equation}
{\omega_\varphi\over\omega_r}={\Delta\varphi_r\over2\pi}=1+q.
\end{equation}

\section{RN Black String}
\label{rn}

We consider the RN black string. It turns out that
\begin{equation}
f(r)=1-{2M\over r}+{Q^2\over r^2},
\end{equation}
is still the solution of the corresponding $5d$ Einstein's equations. Alternatively, we can perceive this metric as a solution arising from the Einstein-Maxwell compactification along the framework of Quasitopological Electromagnetism (QTE) theory\footnote{We thank Adolfo Cisterna for bringing this formalism into our attention.}~\cite{Cisterna:2020rkc, Cisterna:2021ckn}. The horizons are still located at $r_{\pm}=M\pm\sqrt{M^2-Q^2}$, stretching to extra dimension and creating a hypercylindrical topology. Despite the remarkable simplicity in the metric solution, its geodesic phenomenology is not quite that trivial. There is a rich family of orbital solutions, both for timelike and null particles.

%expand the established black string spacetime via incorporation of charge, in which this case by extending the fundamental charged black hole solution, RN ($1-2M/r+Q^2/r^2$), with an extra spatial dimension. For our purpose, we shall then call it the RN black string. The horizons are still located at $r_{\pm}=M\pm\sqrt{M^2-Q^2}$, \textit{stretching} to extra dimension and creating a hypercylindrical topology.

\begin{figure}[!ht]
	\centering
	\begin{tabular}{cc}
		\includegraphics[height=6.5cm]{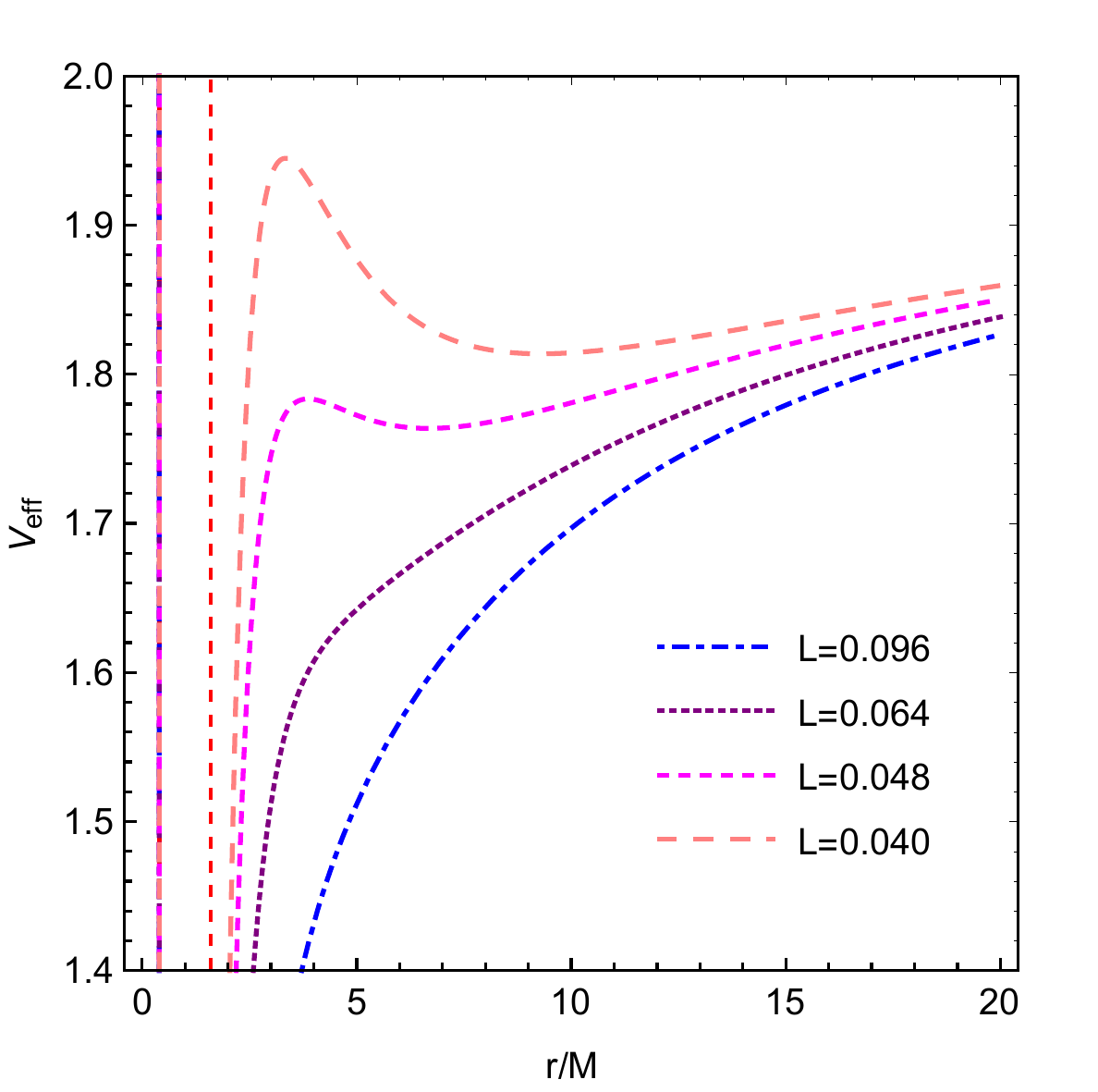} &
		\includegraphics[height=6.5cm]{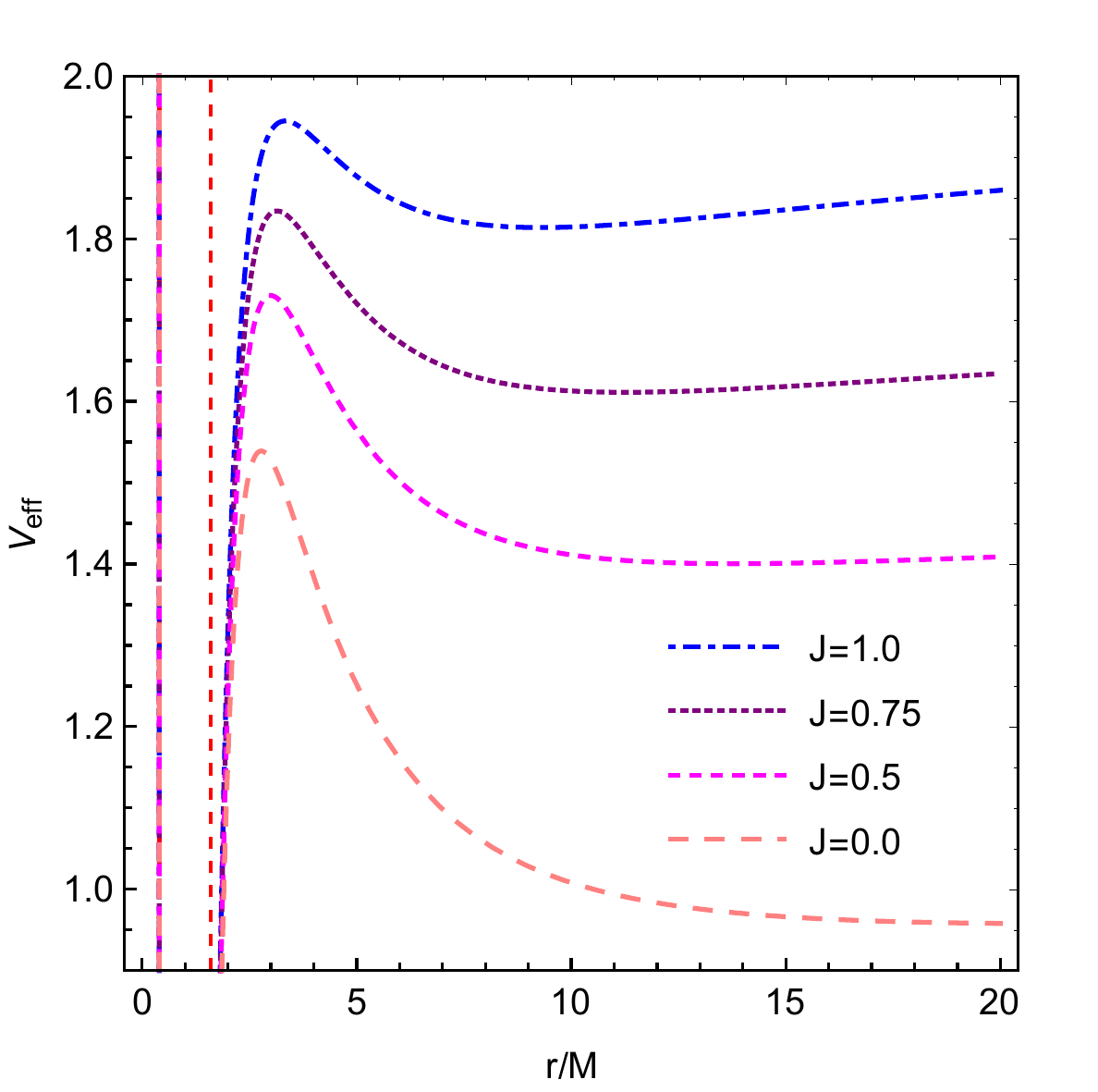}
	\end{tabular}
	\caption{[Left] Plot of effective potential for massive particles in RN two-horizon case with $Q=0.8$ and $J=1$ and various values of $L$ and [Right] similar plot with $L=0.04$ and various values of $J$. The dashed red lines refer to the inner and outer horizons.}
	\label{rn_vt}
\end{figure}

The typical $V_{eff}$ for massive particles is shown on Fig. \ref{rn_vt}. Notice that the local maximum rises along with the increase of $J$ value. Since energy and angular momentum are inversely proportional (see Eq. \eqref{geo}), this means that the existence of bound orbit with bigger value of the constant $J$ requires test particles to have smaller $L$. For massless particles, the $V_{eff}$ is shown on Fig. \ref{rn_vn}. We observe the same trend as revealed in the timelike one. Nonetheless, we notice that in higher angular momentum case, the existence of additional dimension ($J>0$) elevate the potential and create a condition that allows photon to form bound orbits.
\subsection{Timelike CO and ISCO}

The existence of CO and ISCO for timelike geodesic in $4d$ RN black hole has been widely discussed in~\cite{Pugliese:2010ps, Schroven:2020ltb} and the references therein. As stated earlier, the condition for CO, given by Eq.~\eqref{ucosco}, for the RN string translates into
\begin{eqnarray}
\label{MBOmassive}
\left(J-\mathbb{E}^2+1\right) L r_{CO}^4-2 (J+1) M L r_{CO}^3+\left[\left(J+1\right) L Q^2+1\right]r_{CO}^2-2 M r_{CO}+Q^2&=&0,\nonumber\\
2 (J+1)M L r_{CO}^3-2 \left[1-\left(J+1\right) L Q^2\right]r_{CO}^2+6M r_{CO}-4 Q^2&=&0,\nonumber\\
\end{eqnarray}
where $r_{CO}$ is the circular radii. Simple algebra shows that those equations can be solved simultaneously to obtain one-parameter class of solutions parametrized by $r_{CO}$,
\begin{eqnarray}
\label{LJ}
L&=&\frac{3 M r_{CO} -2 Q^2-r_{CO}^2}{(J+1) r_{CO}^2 \left(Q^2-M r_{CO}\right)},\nonumber\\
\mathbb{E}^2&=&\frac{(J+1) \left(-2Mr_{CO}+Q^2+r_{CO}^2\right)^2}{r_{CO}^2 \left(-3Mr_{CO}+2 Q^2+r_{CO}^2\right)}.\nonumber\\
\end{eqnarray}
In the extremal case we have
\begin{equation}
L={r-2M\over \left(1+J\right) M r^2},\ \ \ \mathbb{E}^2={\left(1+J\right)\left(r-M\right)^3\over\left(r-2M\right) r^2}.
\end{equation}
When $J\rightarrow0$ they reduce to the circular motion conditions for neutral test particles around $4d$ RN black hole~\cite{Pugliese:2010ps}.

For any given $L$ the ISCO is the smallest radius of circular orbit before the particle plunges into the black hole. It is the inflection point of $V_{eff}$. From condition~\eqref{isco}, the additional equation to be solved is
\begin{equation}
4 (J+1) L M r_{ISCO}^3-6\left[(J+1) L Q^2+1\right]r_{ISCO}^2 +24 M r_{ISCO}-20 Q^2=0.
\end{equation}
Since the extra-dimensional signature always appears multiplying $L$, {\it i.e.,} $(1+J)L$, then from the expression of $L$ in~\eqref{LJ} it is obvious that the ISCO radii satisfy  
\begin{equation}
Mr_{ISCO}^3-6M^2r_{ISCO}^2+9MQ^2r_{ISCO}-4Q^4=0.
\end{equation}
The same thing happens for the extremal black strings. The ISCO radii becomes
\begin{equation}
r_{ISCO}^2-5Mr_{ISCO}+4M^2=0,
\end{equation}
with solutions
\begin{equation}
r_{ISCO}=\{M,\ 4M\},
\end{equation}
the same as the extremal $4d$ RN ISCO; the timelike ISCO is oblivion to the existence of extra dimension ({\it i.e.,} independent of $J$). Notice that this is precisely the same condition for ISCO in $4d$ RN black hole~\cite{Pugliese:2010ps}.

\subsection{Null Circular Orbit}

For null geodesic ($\epsilon=0$) it is customary to define {\it impact parameter} $b$, where $b^2\equiv 1/\mathbb{E}^2L$. The geodesic equation~\eqref{geo} becomes 
\begin{equation}
\dot{r}^2+V_{eff}=b^2,\ \ \ V_{eff}\equiv f(r)\left(j+{1\over r^2}\right),
\end{equation}
where $j\equiv JL$ and we subsequently rescale the proper time $\tau\rightarrow\tau/L$. It is well-known that $4d$ RN spacetime does not support stable CO, except in the extremal limit. To be precise, for any given charge $Q/M$ the $\partial_rV_{eff}=0$ condition for null gives us a quadratic equation whose roots are 
\begin{equation}
r^{\pm}_{CO}={3M\pm\sqrt{9M^2-8Q^2}\over2}.
\end{equation}
One can easily observe that $r^-_{CO}$ is at the local minima of $V_{eff}$ and is between the two horizons. Thus, the only observable CO, called the {\it photon sphere}, $r^+_{CO}$ is unstable~\cite{Pradhan:2010ws, Khoo:2016xqv}. At $M=Q$, the CO radii shifts further into 
\begin{equation}
\label{4dextr}
r^{\pm}_{CO}=\{2M,\ M\};
\end{equation}
 the stable $r_{CO}$ coincides with the extremal horizon $r_{extr}$.

The situation is rather different in this $5d$ black string. The additional term that multiplies $j$ in $V_{eff}$ gives qubic equation for the $\partial_rV_{eff}=0$,
\begin{equation}
\label{nullco}
j M r_{CO}^3-\left(j Q^2+1\right)r_{CO}^2 +3M r_{CO}-2Q^2=0,
\end{equation}
{\it i.e.,} in general we have an additional local minima of $V_{eff}$. By adjusting $j$ we can have a stable photon sphere outside the outer horizon, $r_{ps}>r_+$. For extremal case, Eq.~\eqref{nullco} has three roots
\begin{equation}
\label{exco}
r_{CO}^{e,\pm}=\bigg\{M,\ {1\pm\sqrt{1-8jM^2}\over 2jM}\bigg\},
\end{equation}
one of which lies on the extremal horizon, $r^e_{CO}=r_{extr}=M$. The other two ($r_{\pm}$) are real and located outside $r_{extr}$ provided the following condition is satisfied,
\begin{equation}
0<j<{1\over8M^2}.
\end{equation}
The $r_{CO}^+$ ($r_{CO}^-$) act as the stable (unstable) radii. As $j\rightarrow0$ we have $r_{CO}^{\pm}\rightarrow\{2M, \infty\}$ and we recover the condition~\eqref{4dextr}.

\begin{figure}[!ht]
	\centering
	\begin{tabular}{cc}
		\includegraphics[height=6.5cm]{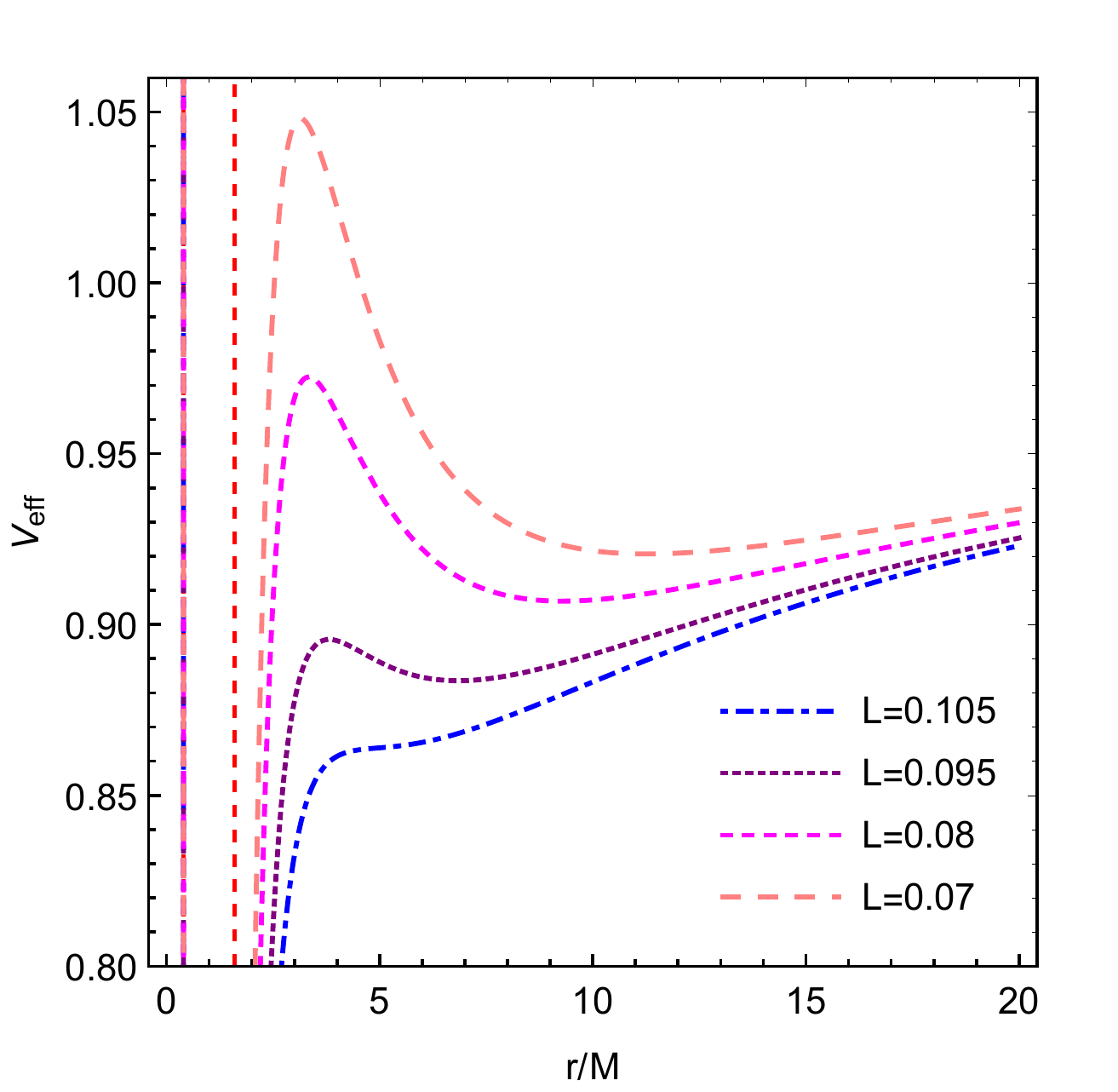} &
		\includegraphics[height=6.5cm]{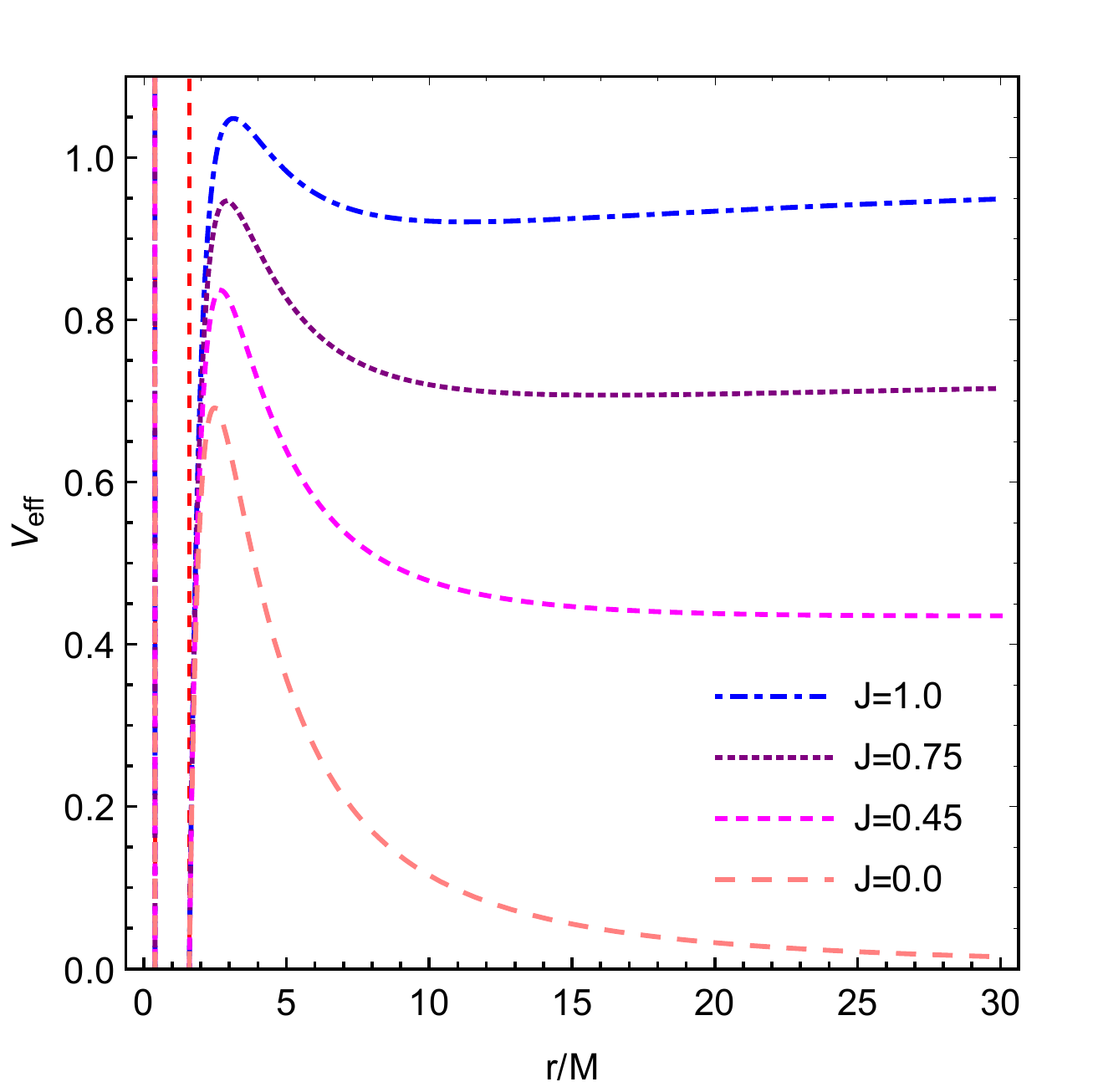}
	\end{tabular}
	\caption{[Left] Plot of effective potential for light particles in RN  two-horizon case with $Q=0.8$ and $J=1$ and various values of $L$ and [Right] similar plot with $L=0.07$ and various values of $J$. The dashed red lines refer to the inner and outer horizons.}
	\label{rn_vn}
\end{figure}

\subsection{Exact Solutions of the Geodesic}

Studying the shape of $V_{eff}$ can only tell us to so much about the qualitative types of orbits. To know their specific shapes we must solve the geodesic equation.
Using the black string metric above and setting $r=r(\phi)$, the geodesic equation~\eqref{geo} can be cast into% it as a function of $\phi$, given as
\begin{equation}
\label{orbitr}
\bigg(\frac{dr}{d\phi}\bigg)^2 = \mathbb{E}^2 L r^4 - f(r) \left(r^2 +  L r^4 (J+\epsilon) \right) \equiv \mathcal{R}(r).
\end{equation}
This can be simplified by rescaling $r\rightarrow r/M$, $Q\rightarrow Q/M$, $L\rightarrow L/M^2$, and defining:\\
$v\equiv L(\mathbb{E}^2-(J+\epsilon )),\ w\equiv2 L (J+\epsilon ),\ x\equiv-L Q^2 (J+\epsilon )-1,\  y\equiv2,$ and $z\equiv-Q^2,$
%\end{eqnarray}
such that Eq. \eqref{orbitr} now reads
\begin{equation}
\label{orbitR}
\bigg(\frac{dr}{d\phi}\bigg)^2 = vr^4 +wr^3 +xr^2 +yr +z.
\end{equation}

Further simplification by expanding $r$ around (one of its zeros) $r=s+r_0$ and defining:
%\begin{equation}
$a_0\equiv v, \ a_1\equiv4 r_0 v+w, \ a_2\equiv6 r_0^2 v + 3 r_0 w + x,$ and $a_3\equiv4 r_0^3 v + 3 r_0^2 w + 2 r_0 x + y$,
%\end{equation}
transform Eq. \eqref{orbitR} into
\begin{equation}
\label{orbitS}
\bigg(\frac{ds}{d\phi}\bigg)^2 = a_0 s^4 +a_1 s^3 +a_2 s^2 +a_3 s.
\end{equation}
The degree of the corresponding polynomial can be reduced by setting $r\equiv1/u$, and further by $u\equiv4 y/a_3-a_2/3 a_3$. We then obtain the Weierstrass form
\begin{eqnarray}
\label{weier}
\bigg(\frac{dy}{d\phi}\bigg)^2 = 4 y^3 -g_2 y - g_3,
\end{eqnarray}
where $g_2$ and $g_3$ are given as
\begin{eqnarray}
\label{g2}
g_2&=&\frac{1}{16} \left(\frac{4 a_2^2}{3}-4 a_1 a_3 \right), \\
\label{g3}
g_3&=&\frac{1}{16} \left(\frac{1}{3} a_1 a_3 a_2-\frac{1}{27} 2 a_2^3-a_0 a_3^2\right).
\end{eqnarray}

Eq.~\eqref{orbitr} admits analytical solution in terms of Weierstrass $\wp$-function~\cite{Hackmann:2010tqa}:
\begin{equation}
\label{rsol}
r(\phi) = r_0 + \frac{3 a_3}{12\wp(\phi-\phi_{in}) - a_2}
\end{equation}
where the initial angle $\phi_{in}$ is only dependent on $\phi_0$ and $y_0$:
\begin{equation}
\label{rint}
\phi_{in} = \phi_0 +\int_{y_0}^{\infty} \frac{dy}{\sqrt{ 4 y^3 -g_2 y - g_3}}, \,\, \ \ \ \ y_0\equiv\frac{1}{4} \bigg( \frac{a_3}{r_{in}-r_0}+\frac{a_2}{3} \bigg).
\end{equation}
Making use of \eqref{tdot} we have the equation for $\omega$ as
\begin{equation}
\label{orbitw}
\bigg( \frac{d\omega}{d\phi} \bigg) = J \sqrt{L} r(\phi)^2,
\end{equation}
which gives the solution of $\omega$ in terms of $\phi$ as
\begin{equation}
\label{wsol}
\omega = J \sqrt{L} \int_{\phi_0}^{\infty} r(\phi)^2 d\phi.
\end{equation}
Eq.~\eqref{wsol} is later to play a significant role in enabling null bound orbit outside the string's horizons.

\begin{figure}[!ht]
	\centering
	\begin{tabular}{cc}
		\includegraphics[height=12cm]{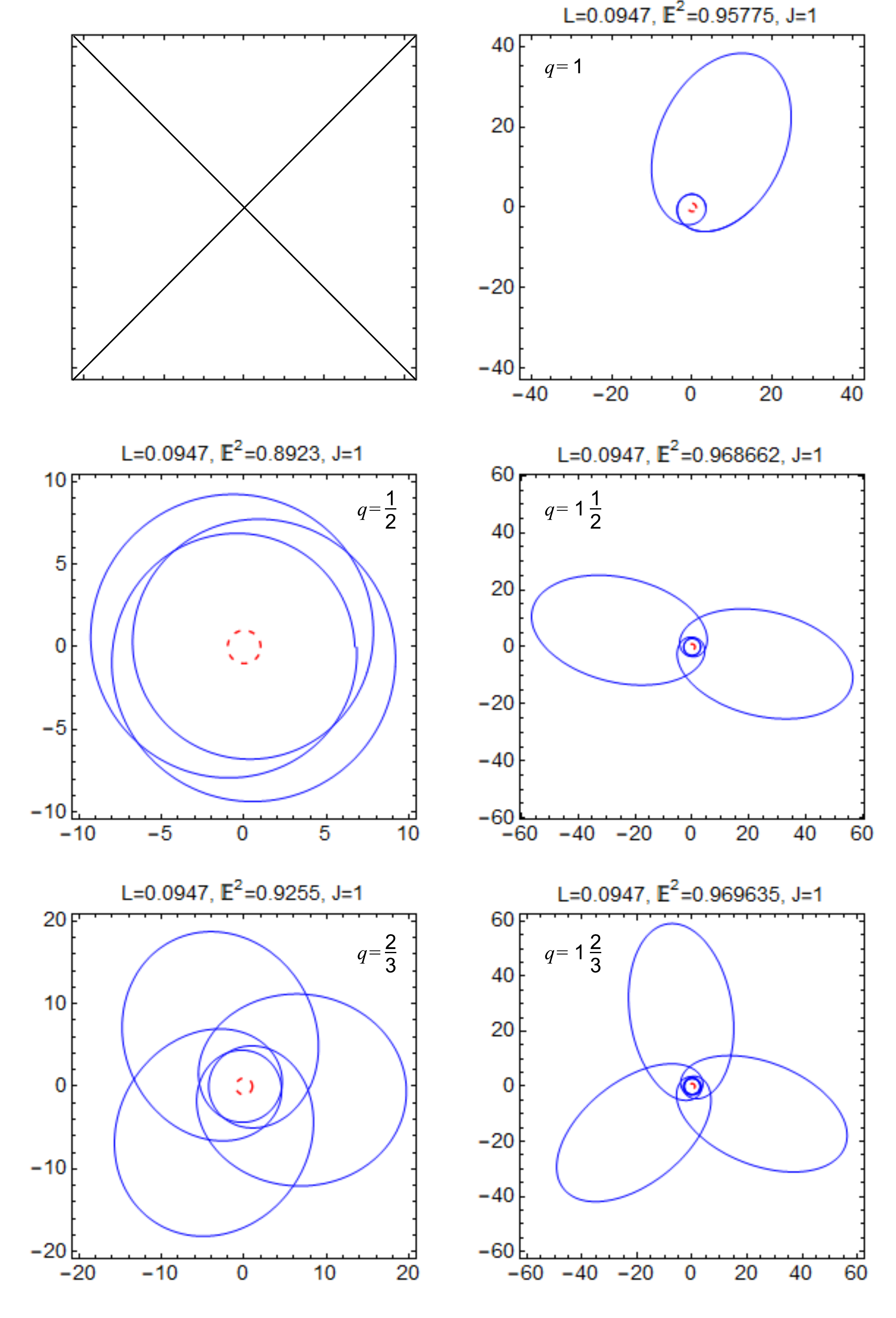}&
		\includegraphics[height=12cm]{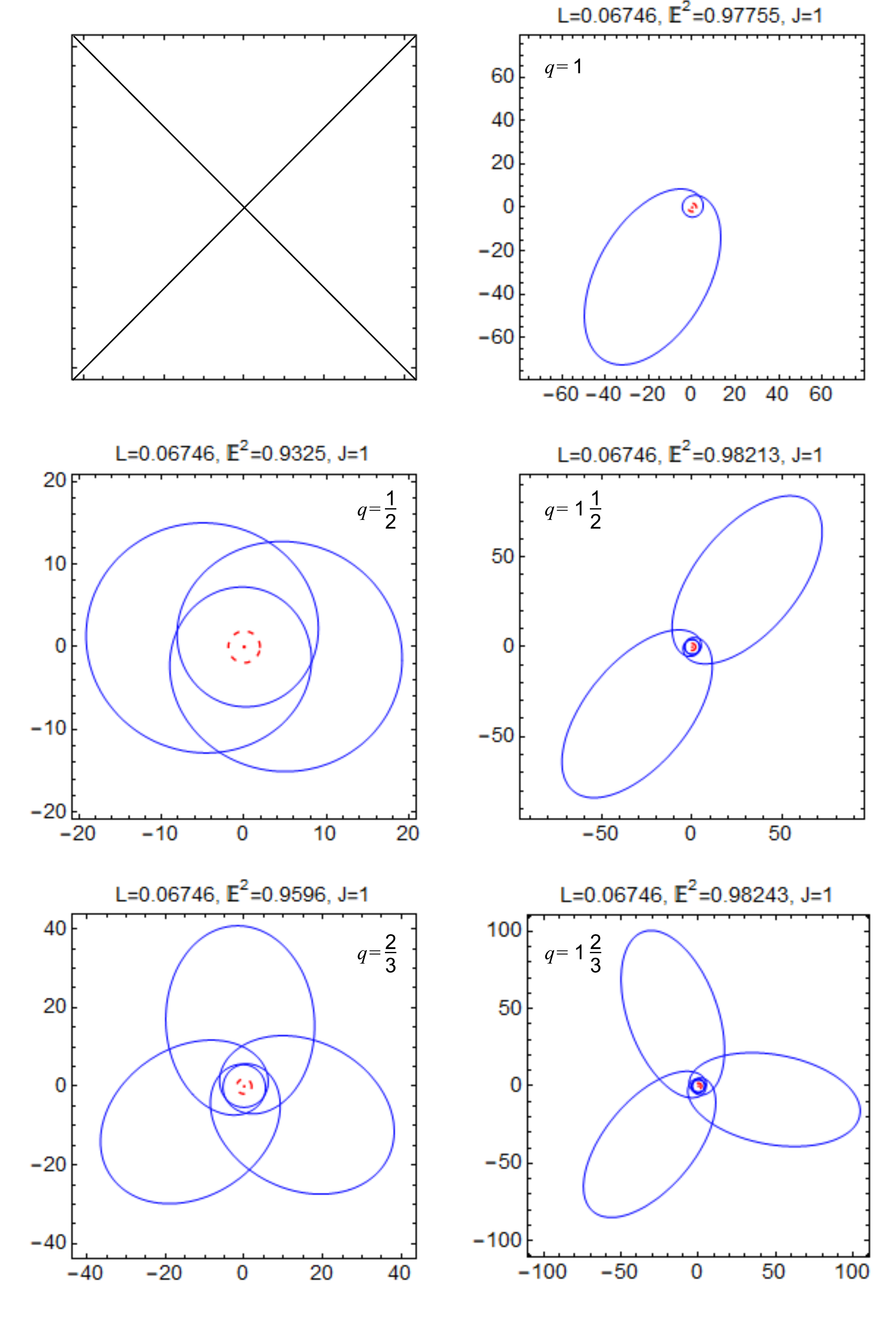}
	\end{tabular}			
	\caption{The $z=1,2,3$ orbits in RN null condition: extremal case with $w=0$ for the first column and $w=1$ for the second column; two-horizon case with $w=0$ for the third column and $w=1$ for the fourth column. Note that the first entries in the first and third column are blank because the $q=0+\frac{0}{1}$ orbits are inaccessible.}
	\label{rn_n}
\end{figure}
\begin{figure}[!ht]
	\centering
	\begin{tabular}{cc}
		\includegraphics[height=12cm]{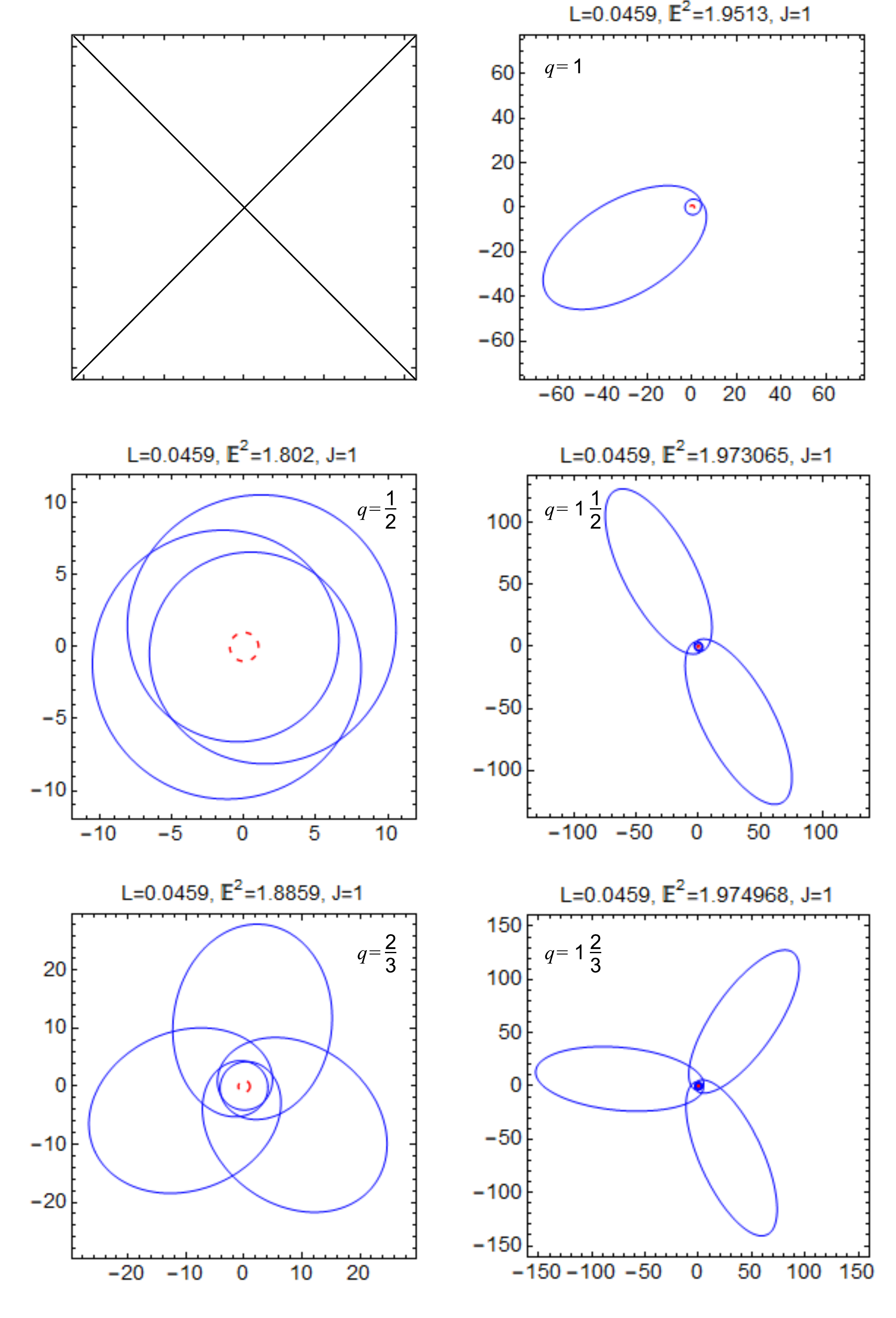}&
		\includegraphics[height=12cm]{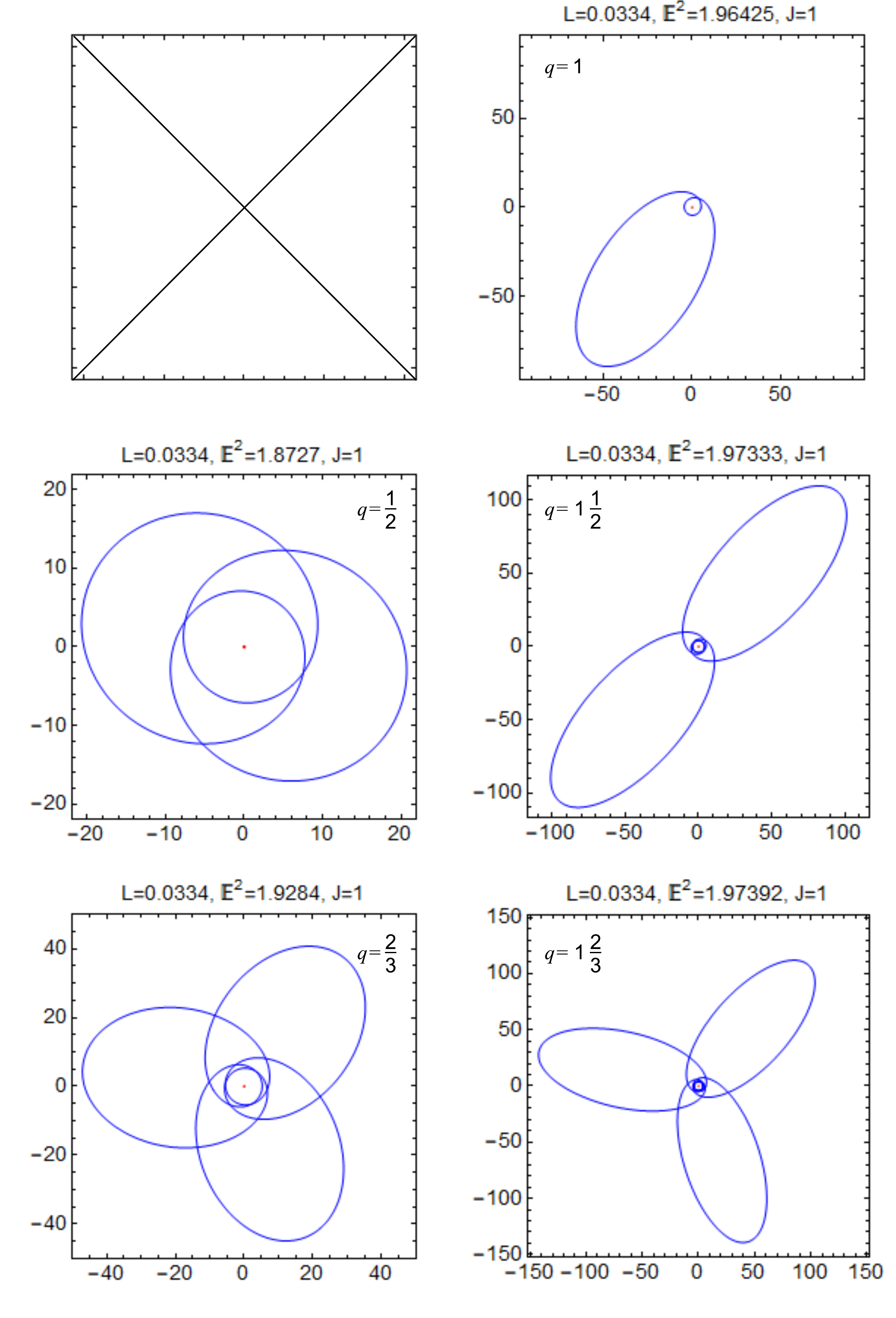}
	\end{tabular}			
	\caption{The $z=1,2,3$ orbits in RN timelike condition: extremal case with $w=0$ for the first column and $w=1$ for the second column; two-horizon case with $w=0$ for the third column and $w=1$ for the fourth column. Note that the first entries in the first and third column are blank because the $q=0+\frac{0}{1}$ orbits are inaccessible.}
	\label{rn_t}	
\end{figure}

\subsection{RN Black String's Bound Orbit Taxonomy}
\label{rn_taxo}

%As in the case of RN black hole, RN black string can be extremal ($M^2=Q^2$), non-extremal ($M^2>Q^2$), or naked ($M^2<Q^2$). 

As discussed above, the black string's $V_{eff}$ is quartic in $r$, so in general it gives three local extrema; an additional local minimum outside the horizon. In this work we limit ourselves only to the case of physical orbits; that is, which lies outside the  horizon and can thus be observable. We avoid discussing the tricky case of horizon-crossing orbits despite the premise of many-world orbits by Grunau and Kagramanova, in which the orbit is viewed as infinite continuity of the patches of the spacetime \cite{Grunau:2010gd}. Armed with the aforementioned taxonomy in \ref{taxo}, we carefully examine the bounded domain of region IV (Fig.~\ref{region}) in RN black string potential and present the periodic orbits up to 3 zooms and 2 whirls. The whole existence of stable null orbits in our case is genuine. The $J$-factor leads to the existence of periodic orbits, as opposed to the $4d$ case where no such stable orbits present~\cite{Misra:2010pu, Pradhan:2010ws}. The plots are shown in Fig. \ref{rn_n}.

\begin{table}[!h]
	\setlength{\tabcolsep}{0.8em}
	\begin{tabular}{cccc}
		\begin{tabular}{||c |} 
			\hline
			$q$ \\ \hline
			SCO \\ \hline
			1/2 \\ \hline
			2/3 \\ \hline
			1     \\ \hline
			1 1/2 \\ \hline
			1 2/3 \\ \hline
			UCO \\ \hline
		\end{tabular}
		\begin{tabular}{|c | c |} 
			\hline
			$\mathbb{E}$ (Extremal) & Position \\ \hline
			1.791566220 & 0.00\% \\ \hline
			1.802000000 & 5.62\% \\ \hline
			1.885900000 & 50.79\% \\ \hline
			1.951300000 & 86.00\% \\ \hline
			1.973065000 & 97.72\% \\ \hline
			1.974968000 & 98.75\% \\ \hline
			1.977293000 & 100.00\% \\ \hline
		\end{tabular}
		\begin{tabular}{|c | c |} 
			\hline
			$\mathbb{E}$ (Two-horizon) & Position \\ \hline
			1.841640600 & 0.00\% \\ \hline
			1.872700000 & 23.38\% \\ \hline
			1.928400000 & 65.31\% \\ \hline
			1.964250000 & 92.30\% \\ \hline
			1.973330000 & 99.14\% \\ \hline
			1.973920000 & 99.58\% \\ \hline
			1.974477000 & 100.00\% \\ \hline
		\end{tabular}
		\begin{tabular}{|c ||} 
			\hline
			Deviation \\ \hline
			- \\ \hline
			12.56\% \\ \hline
			10.27\% \\ \hline
			4.45\% \\ \hline
			1.00\% \\ \hline
			0.59\% \\ \hline
			- \\ \hline
		\end{tabular}
	\end{tabular}
	\caption{Comparison of energy level and the position of particular orbit inside the energy range as percentage for timelike case in extremal and two-horizon RN black string. SCO refers to the bottom level of energy range (no whirl), while UCO refers to the top one (maximum whirl).}
	\label{rn_t12}
\end{table}
\begin{table}[!hb]
	\setlength{\tabcolsep}{0.8em}
	\begin{tabular}{cccc}
		\begin{tabular}{||c |} 
			\hline
			$q$ \\ \hline
			SCO \\ \hline
			1/2 \\ \hline
			2/3 \\ \hline
			1     \\ \hline
			1 1/2 \\ \hline
			1 2/3 \\ \hline
			UCO \\ \hline
		\end{tabular}
		\begin{tabular}{|c | c |} 
			\hline
			$\mathbb{E}$ (Extremal) & Position \\ \hline
			0.891967985 & 0.00\% \\ \hline
			0.894300000 & 2.96\% \\ \hline
			0.925500000 & 42.51\% \\ \hline
			0.957750000 & 83.39\% \\ \hline
			0.968662000 & 97.22\% \\ \hline
			0.969635000 & 98.46\% \\ \hline
			0.970851807 & 100.00\% \\ \hline
		\end{tabular}
		\begin{tabular}{|c | c |} 
			\hline
			$\mathbb{E}$ (Two-horizon) & Position \\ \hline
			0.919894000 & 0.00\% \\ \hline
			0.932500000 & 20.07\% \\ \hline
			0.959600000 & 63.20\% \\ \hline
			0.977550000 & 91.77\% \\ \hline
			0.982130000 & 99.06\% \\ \hline
			0.982430000 & 99.54\% \\ \hline
			0.982718600 & 100.00\% \\ \hline
		\end{tabular}
		\begin{tabular}{|c ||} 
			\hline
			Deviation \\ \hline
			- \\ \hline
			12.10\% \\ \hline
			14.63\% \\ \hline
			5.93\% \\ \hline
			1.30\% \\ \hline
			0.77\% \\ \hline
			- \\ \hline
		\end{tabular}
	\end{tabular}
	\caption{Comparison of energy level and the position of particular orbit inside the energy range as percentage for null case in extremal and two-horizon RN black string. SCO refers to the bottom level of energy range (no whirl), while UCO refers to the top one (maximum whirl).}
	\label{rn_n12}
\end{table}

Levin in~\cite{Levin:2009sk} shows that, since rational numbers are discrete so are the energy levels that produce the orbits. The existence of bound orbits are bounded by UCO and SCO, so we can quantify the level energy by the deviation from them. We define the corresponding deviation $\eta$ as
\begin{equation}
\eta\equiv{\left(\mathbb{E}-\mathbb{E}_{SCO}\right)\over\left(\mathbb{E}_{UCO}-\mathbb{E}_{SCO}\right)}\bigg|_{timelike/null}.
\end{equation}
In Table \ref{rn_t12} we list the energy level of various periodic orbits according to their energy level, for massive particles, in both extremal and two-horizon RN black string. Similar table for light particles is shown in Table \ref{rn_n12}. % The deviation ($\eta$) refers to the {\it position} (of extremal and two-horizon orbit), defined as the following ratio: 
The table shows increasing number of rational number as it goes down the list. One behavior that is observed in both cases is the standard deviations, while fluctuate, tends to decrease as energy level rises to UCO. If we calculate the average of deviation for timelike and null condition, we obtain the value $6.36\%$ and $6.94\%$, respectively. The reasoning of why we separate the table by horizon case becomes clear when we pay attention to each deviation average: $2.03\%$ for extremal case and $0.86\%$ for two horizon one, which is way smaller than the numbers earlier. This finding means that the distance between energy levels is more consistent in each specific horizon case.

\begin{table}[!ht]
	\setlength{\tabcolsep}{1.5em}
	\centering
	\begin{tabular}{||c||c|c||c|c||}
		\hline
		q & $\Delta\varphi_{r-4d}$  & $\Delta\varphi_{r-5d}$  \\ \hline %$\Delta\varphi_{r-B1}$ & $\Delta\varphi_{r-B2}$ \\ \hline
		1/2 & 31.1018 & 44.0076  \\ \hline %49.8285 & 50.3815 \\ \hline
		2/3 & 34.5575 & 48.8576  \\ \hline%54.0185 & 55.5692 \\ \hline
		1 & 41.4694 & 58.634 \\ \hline%64.715 & 66.6127 \\ \hline
		1 1/2 & 51.8373 & 73.2913  \\ \hline%81.0931 & 83.3911 \\ \hline
		1 2/3 & 48.085 & 78.1832  \\ \hline%85.7237  & 87.6954 \\ \hline
	\end{tabular}
	\caption{Accumulated radial angle for extremal RN black holes (second column) and black strings (third column) strings as a function of $q$.}
\label{accRNext}	
\end{table}

\begin{table}[!ht]
	\setlength{\tabcolsep}{1.5em}
	\centering
	\begin{tabular}{||c||c|c||c|c||}
		\hline
		q & $\Delta\varphi_{r-4d}$ & $\Delta\varphi_{r-5d}$  \\ \hline %$\Delta\varphi_{r-B1}$ & $\Delta\varphi_{r-B2}$ \\ \hline
		1/2 & 35.8142 & 51.5273  \\ \hline %49.8285 & 50.3815 \\ \hline
		2/3 & 39.7936 & 57.2777  \\ \hline%54.0185 & 55.5692 \\ \hline
		1 & 47.7531 & 68.7163  \\ \hline%64.715 & 66.6127 \\ \hline
		1 1/2 & 59.6837 & 85.9175  \\ \hline%81.0931 & 83.3911 \\ \hline
		1 2/3 & 63.6605 & 91.6488  \\ \hline%85.7237  & 87.6954 \\ \hline
	\end{tabular}
	\caption{Accumulated radial angle for non-extremal RN black holes (second column) and black strings (third column) strings as a function of $q$.}
\label{accRNnext}	
\end{table}

For the timelike case we obtain a qualitatively similar pattern of orbits found in RN black hole "zoo" (see~\cite{Misra:2010pu}), as shown in Fig.~\ref{rn_t}. One might also notice that the whirl is stronger near the radius of UCO ($r_{UCO}$), which is the same behaviour observed in the $4d$ RN. The significant difference is that the test particle in RN black string requires higher energy for the orbit to exist ($\mathbb{E} \approx 2$). We can also calculate the accumulated radial angle of black strings' rational orbit for any given $q$. The Eq.~\eqref{radangle} reads
\begin{equation}
\Delta\varphi_r= 2\int_{r_p}^{r_a} \frac{dr}{Lr^2\sqrt{\mathbb{E}^2-\left(1-{2M\over r}+{Q^2\over r^2}\right)\left(\epsilon +J + \frac{1}{L r^2} \right)}}.
\end{equation}
This equation can be integrated numerically. In Table~\ref{accRNext} and~\ref{accRNnext} we show the accumulated angle for extremal and two-horizon RN black strings, respectively. For each case we compare their values with their corresponding $4d$ counterpart from~\cite{Misra:2010pu}. The accumulated radial angle for black strings is higher than for black holes.

%\clearpage

%\clearpage 

\section{Nonsingular Black String}
\label{b}

\begin{figure}[!ht]
	\centering
	\includegraphics[height=6.5cm]{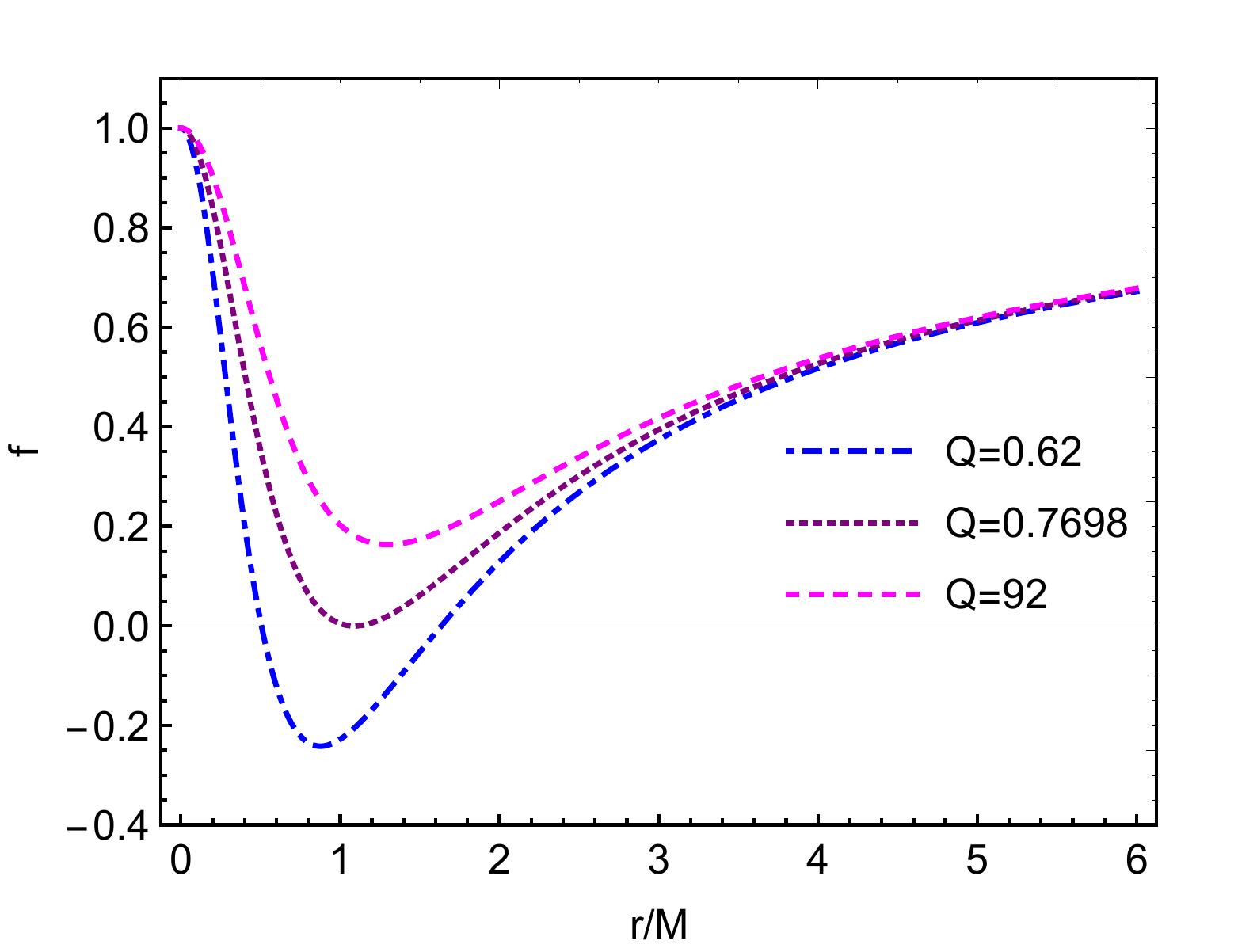} 
	\caption{The three conditions of metric function of Bardeen spacetime.}
	\label{b_f}
\end{figure}

\begin{figure}[!hb]
	\centering
	\begin{tabular}{cc}
		\includegraphics[height=6.5cm]{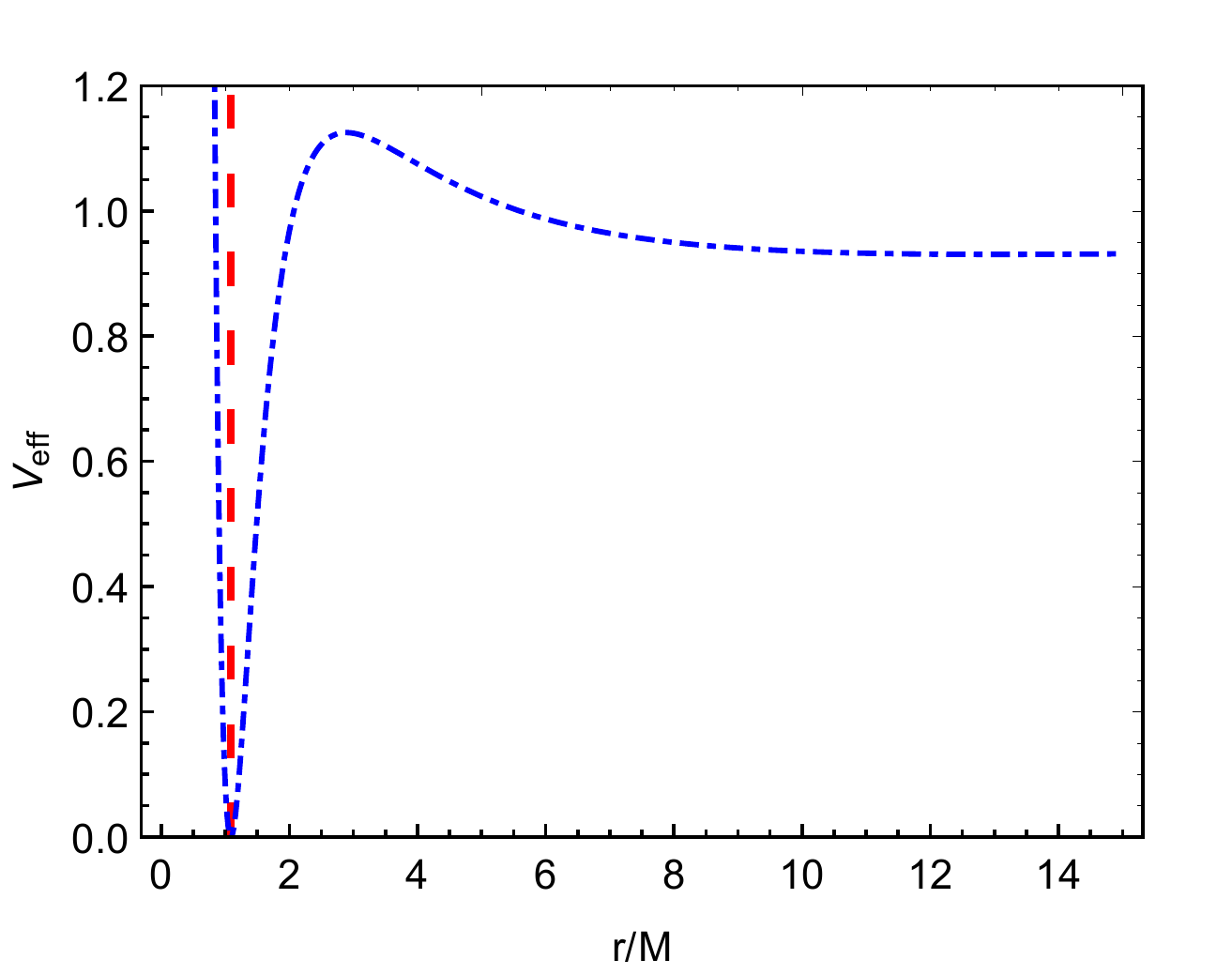} &
		\includegraphics[height=6.5cm]{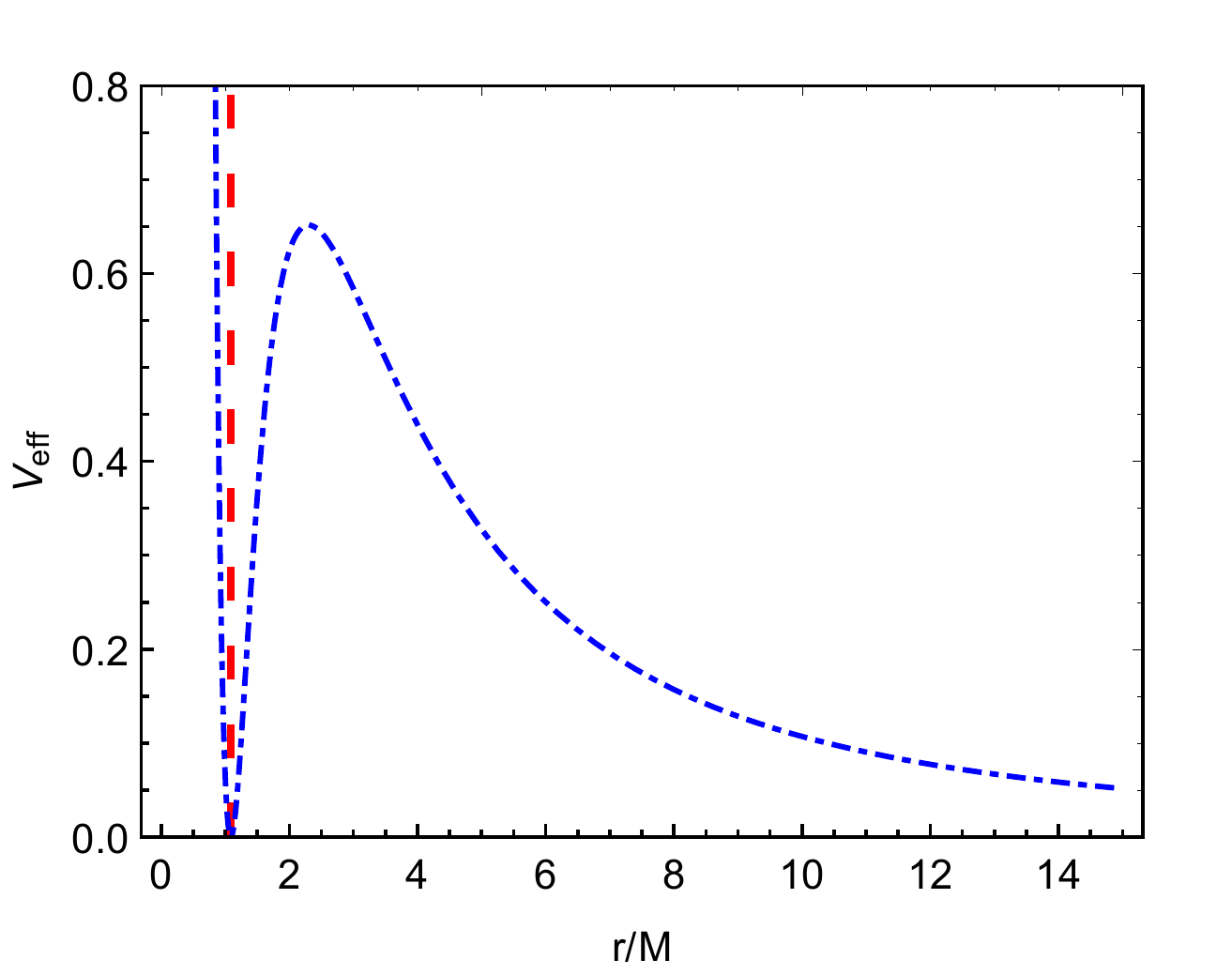}
	\end{tabular}
	\caption{Plot of effective potential in timelike (left) and null geodesics (right) in the extremal $4d$ Bardeen black hole. The dashed red line refers to their respective event horizon.}
	\label{b_vo}
\end{figure}

In 1968 in his seminal paper, Bardeen proposed an analytic solution of a charged BH with no singularity everywhere; a regular\footnote{Many nonsingular black hole models have been developed ever since. For example, the static cases may be found in~\cite{Hayward:2005gi, Bogojevic:1998ma}, while the rotating cases are in~\cite{Bambi:2013ufa,Ghosh:2014hea,Toshmatov:2014nya,Dymnikova:2015hka,Abdujabbarov:2016hnw}.} BH~\cite{bardeen}. The Bardeen solution is given by
\begin{equation}
\label{abg_sol}
f(r)=1-\frac{2 M r^2}{(r^2 + Q^2)^{3/2}}, 
\end{equation}
where $Q$ can be identified as a charge. Surely this solution is different from RN, but also reduces to Schwarzschild in the limit of vanishing $Q$. The Kretschmann scalar is finite everywhere, including at the center
\begin{equation}
\lim_{r\rightarrow0}R^{\alpha\beta\gamma\delta}R_{\alpha\beta\gamma\delta}={96m^2\over Q^{8/3}}.
\end{equation}
At $r\rightarrow0$ the spacetime behaves de Sitter-like, $f(r)\approx1-2mr^2/Q^3$.
%\end{equation}
The horizons $r_h$ are given as the roots of
\begin{equation}
\left(r_h^2+Q^2\right)^3-4 M^2 r_h^4=0.
\end{equation}
In general there are at most two horizons. The typical values of $Q$ for each horizon condition are shown in Fig. \ref{b_f}. The extremal Bardeen is obtained when~\cite{Ayon-Beato:2000mjt}:
\begin{equation}
Q^2={16\over27}M^2\ \longrightarrow r_{extr}=\sqrt{32\over27}M,
\end{equation}
beyond which the black hole becomes naked. This no-horizon case does not violate the cosmic censorship theorem~\cite{Penrose:1964wq} since no singularity is present, and indeed the formation of naked Bardeen spacetime as a result of destroying its event horizon is strongly supported theoretically~\cite{Li:2013sea}. As in the RN case, Bardeen lacks stable null bound orbit outside the horizons and the SCO exists only in the extremal case, where $r_{SCO}=r_{extr}$. This can be seen in Fig. \ref{b_vo}.% Note that since there is no singularity the many-world orbits in Bardeen is not as problematics as in RN.

\begin{figure}[!ht]
	\centering
	\begin{tabular}{cc}
		\includegraphics[height=6.5cm]{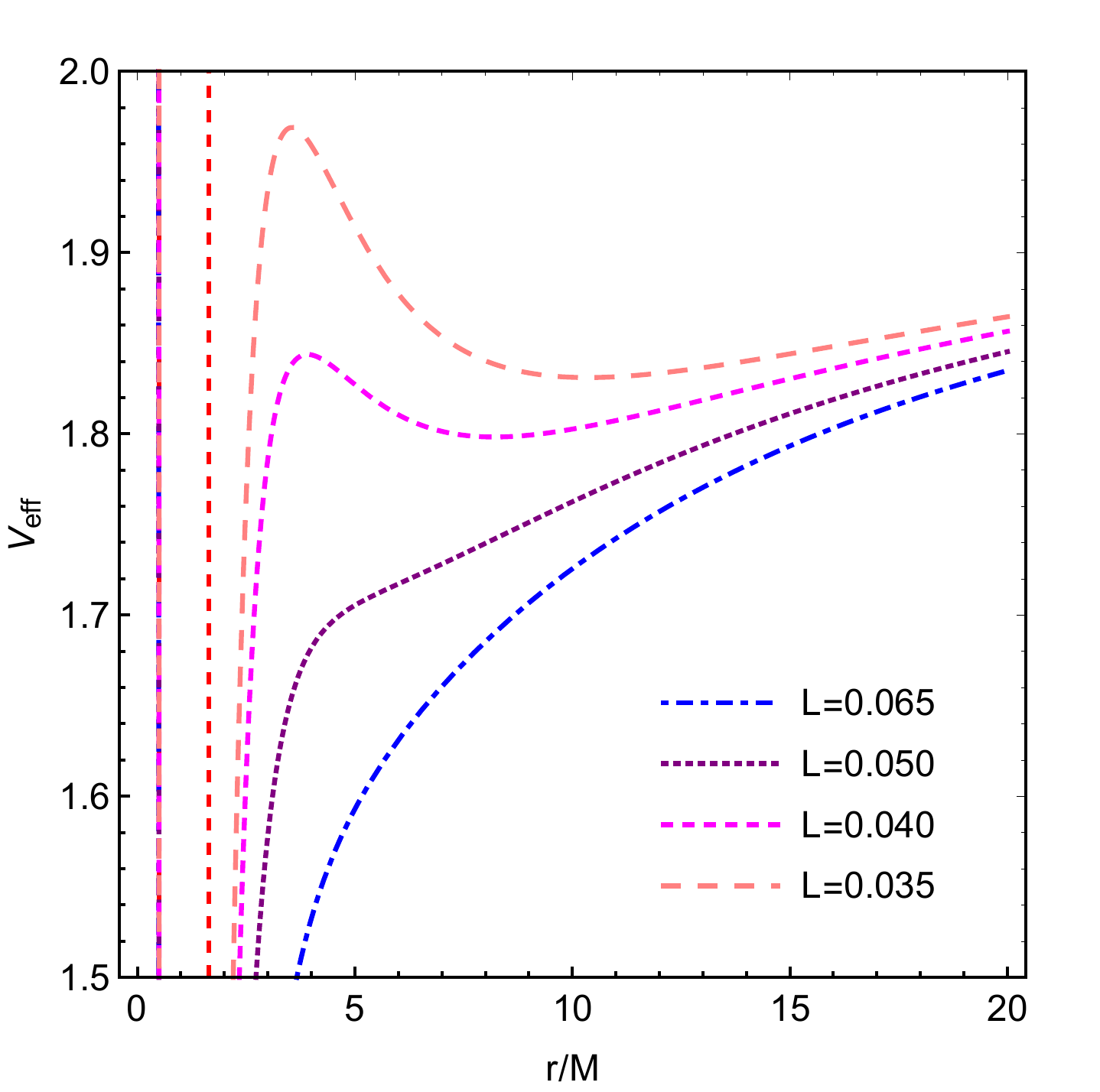} &
		\includegraphics[height=6.5cm]{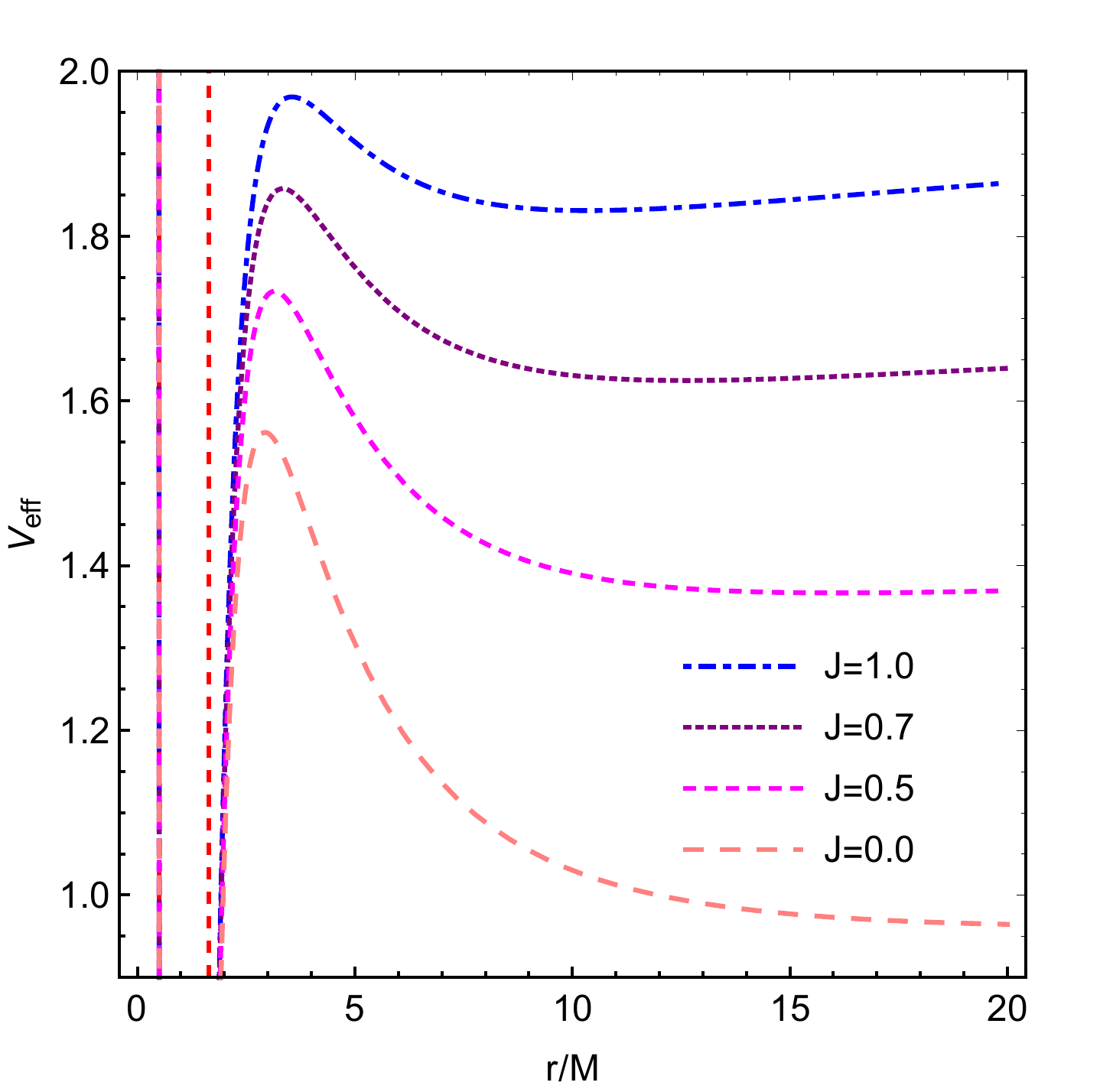}
	\end{tabular}
	\caption{[Left] Plot of effective potential for massive particles in $5d$ two-horizon case with $Q=0.62$ and $J=1$ and various values of $L$ and [Right] similar plot with $L=0.035$ and various values of $J$. The dashed red lines refer to the inner and outer horizons.}
	\label{b_vt}
\end{figure}

\begin{figure}[!ht]
	\centering
	\begin{tabular}{cc}
		\includegraphics[height=6.5cm]{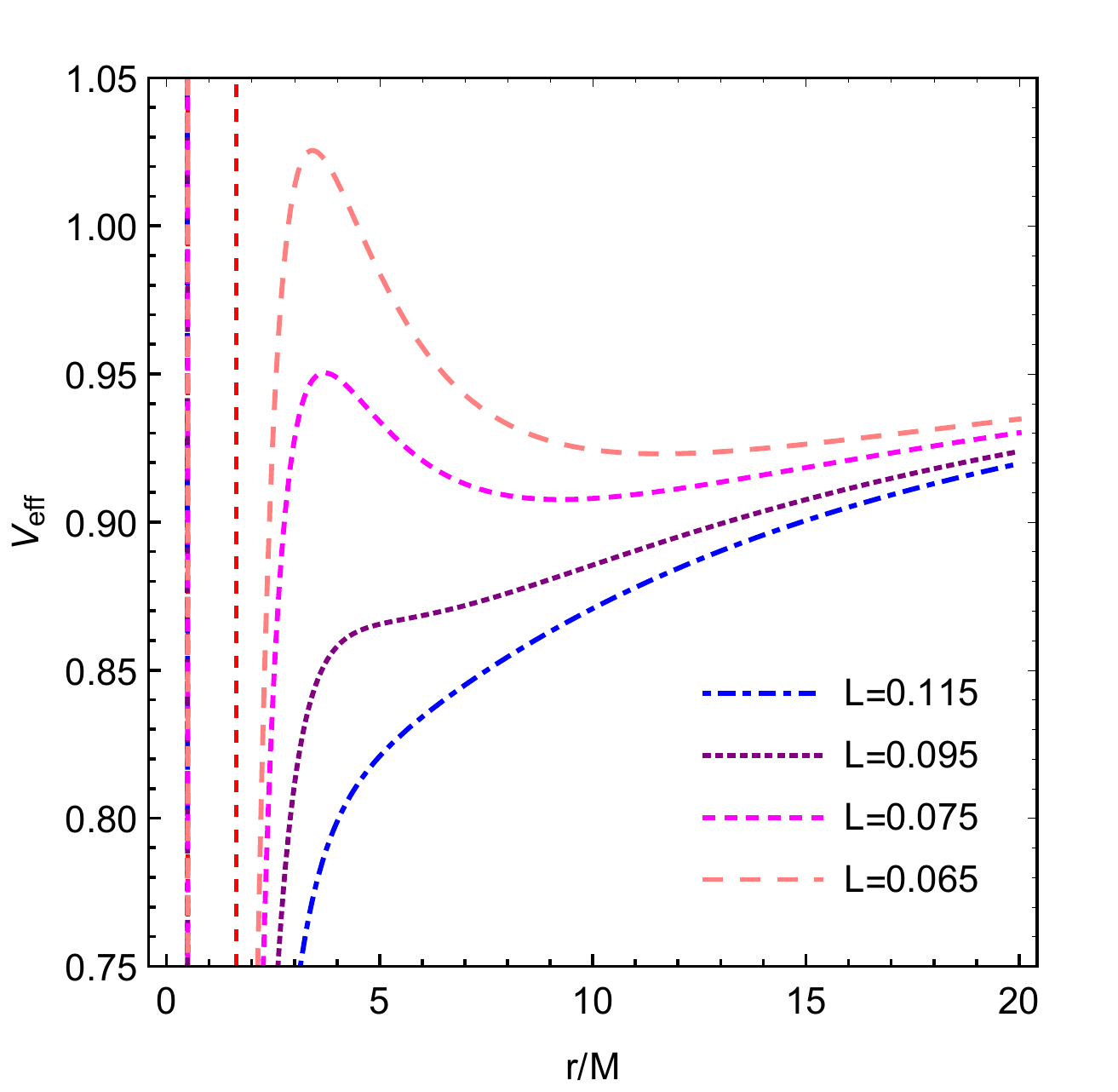} &
		\includegraphics[height=6.5cm]{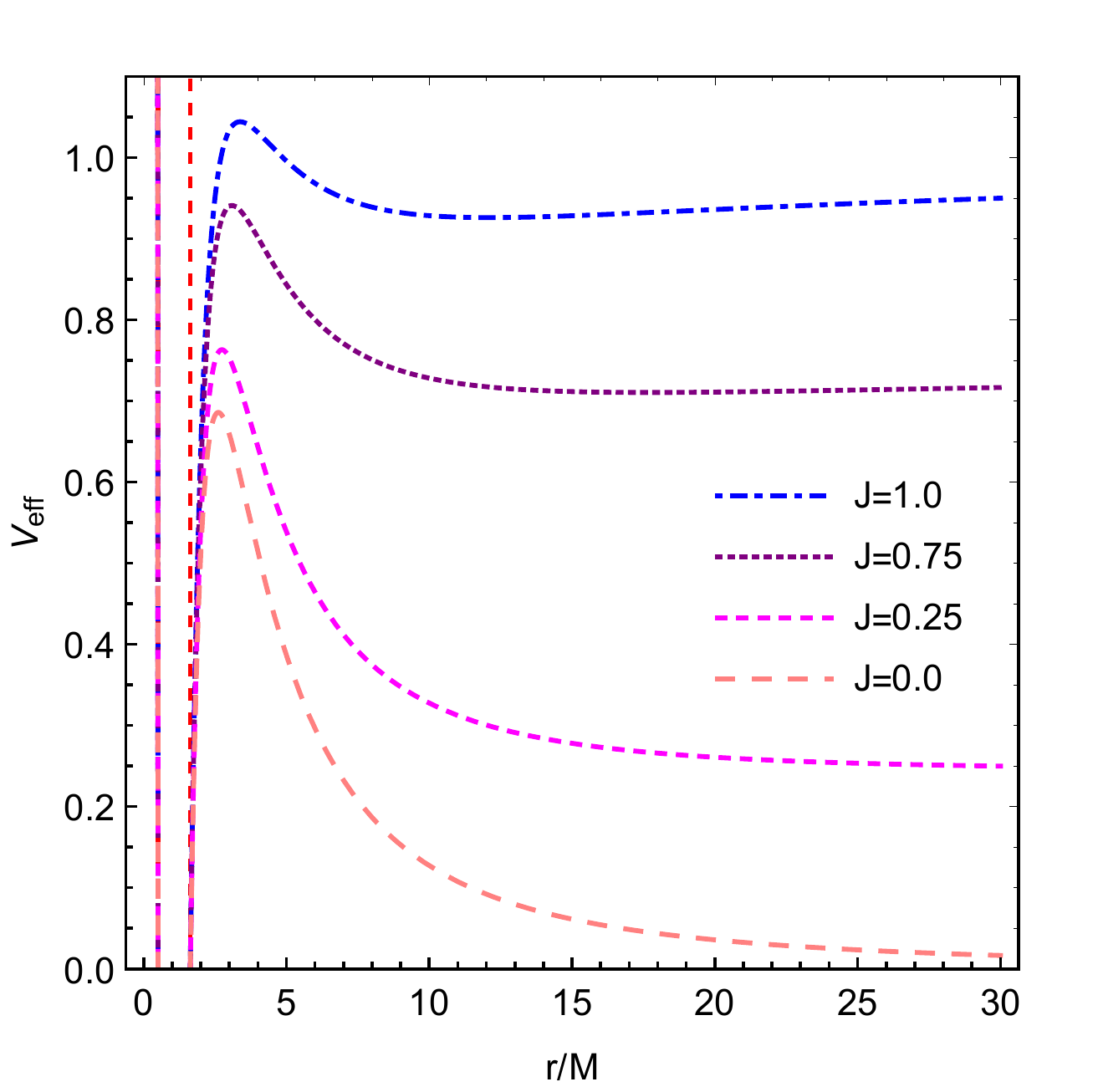}
	\end{tabular}
	\caption{[Left] Plot of effective potential for null particles in $5d$ two-horizon case with $Q=0.62$ and $J=1$ and various values of $L$ and [Right] similar plot with $L=0.063$ and various values of $J$. The dashed red lines refer to the inner and outer horizons.}
	\label{b_vn}
\end{figure}

When extended to $5d$ the metric~\eqref{abg_sol} describes a nonsingular black string. The appearance of $J$ creates a new dynamics different from its $4d$ counterpart, as can be seen in Fig.~\ref{b_vt} and~\ref{b_vn}. As in the RN case, we shall analyze the circular (timelike and null) orbits before obtaining the geodesic's exact solutions.

\subsection{Timelike Circular Orbit}

Applying the condition~\eqref{ucosco} for the black string yields
\begin{eqnarray}
\label{BardeenCO}
\left(J+1-\mathbb{E}^2\right)Lr_{CO}^2\left(r_{CO}^2+Q^2\right)^{3/2}-2M\left(J+1\right)Lr_{CO}^4+\left(r_{CO}^2+Q^2\right)^{3/2}-2Mr_{CO}^2&=&0,\nonumber\\
\left(J+1\right)MLr_{CO}^6-\left(r_{CO}^2+Q^2\right)^{5/2}+\left[3+2\left(J+1\right)LQ\right]Mr_{CO}^4&=&0.\nonumber\\
\end{eqnarray}
One can easily verify that upon $J\rightarrow0$ they reduce to the MBO conditions for $4d$ Bardeen black hole~\cite{Gao:2020wjz}. Solving them for $L$ and $\mathbb{E}^2$ yield
\begin{eqnarray}
L&=&\frac{\left(Q^2+r_{CO}^2\right)^{5/2}-3 M r_{CO}^4}{(J+1) M r_{CO}^4 \left(r_{CO}^2-2 Q^2\right)},\nonumber\\
\mathbb{E}^2&=&\frac{(J+1) \left(Q^2 \left(3 r_{CO}^4-4 M r_{CO}^2 A\right)+4 M r_{CO}^4 \left(M-A\right)+Q^6+3
   Q^4 r_{CO}^2+r_{CO}^6\right)}{-3 M r_{CO}^4 A+Q^6+3 Q^4 r_{CO}^2+3 Q^2 r_{CO}^4+r_{CO}^6},\nonumber\\
\end{eqnarray}
where $A\equiv\sqrt{Q^2+r_{CO}^2}$. 

The ISCO condition gives the additional condition:
\begin{eqnarray}
3\left(Q^2+r_{ISCO}^2\right)^{7/2}+M r_{ISCO}^4 \bigg[-2 (J+1) L Q^4+Q^2 \left(11 (J+1) L r_{ISCO}^2+3\right)\nonumber\\
-2 r_{ISCO}^2 \left((J+1) L r_{ISCO}^2+6\right)\bigg]=0.\nonumber\\
\end{eqnarray}
In the extremal case, $Q^2=16/27$, the condition becomes
\begin{eqnarray}
2 (J+1) L M r_{ISCO}^8+2M\left[3\frac{88}{27} (J+1)LM^2\right]r_{ISCO}^6+{16M^3\over9}\left[\frac{32}{81} (J+1)LM^2-5\right]r_{ISCO}^4\nonumber\\
-\frac{512 M^5r_{ISCO}^2}{243}+3 \left(\frac{16 M^2}{27}+r_{ISCO}^2\right)^2 \left[2 M r_{ISCO}^2-\left(\frac{16
   M^2}{27}+r_{ISCO}^2\right)^{3/2}\right]=0,\nonumber\\
\end{eqnarray}
which can be determined numerically.

\subsection{Null Circular Orbit}

Null geodesic in $4d$ Bardeen spacetime has been discussed widely, for example in~\cite{Eiroa:2010wm, He:2021htq}. In $5d$ by defining $b^2$ and $j$, and by rescaling the proper time the effective potential can be written as
\begin{equation}
V_{eff}=\left(1-\frac{2 M r^2}{(r^2 + Q^2)^{3/2}}\right)\left(j+{1\over r^2}\right).
\end{equation}
One can see that the appearance of $j$-term implies the existence of another turning point in $V_{eff}$ outside the horizon, as can be seen in Fig.~\ref{b_vn}. The CO radii are determined by the roots of the following equation
\begin{equation}
2 M r_{CO}^4 \left(j \left(r_{CO}^2-2 Q^2\right)+3\right)-2 \left(Q^2+r_{CO}^2\right)^{5/2}=0.
\end{equation}
For the extremal case, the condition becomes
\begin{eqnarray}
-81 \sqrt{3} M r_{CO}^4 \left(J L \left(32 M^2-27 r_{CO}^2\right)-81\right)-\left(16 M^2+27 r_{CO}^2\right)^{5/2}=0.\nonumber\\
\end{eqnarray}
Both conditions are polynomial equations of degree six, and the roots can best be found numerically.

\subsection{Exact Solutions and the Taxonomy of Bound Orbits}

%It is not known how Bardeen got his solution, or what matter sources such a regular spacetime. Recently there have been attempts to explain Bardeen BH as being sourced by NLED electric as well as magnetic charge~\cite{Stuchlik:2014qja, Ayon-Beato:1998hmi, Ayon-Beato:1999kuh, Ayon-Beato:2000mjt, Rodrigues:2018bdc}. This nonlinearity and distinctness from RN is what makes Bardeen regular. The timelike geodesics and its corresponding phenomenology around this NLED-Bardeen were studied, for example, in~\cite{Garcia:2013zud, Ghaffarnejad:2016dlw, Vrba:2019vqh, Rayimbaev:2020hjs}. It was recently pointed out in~\cite{Habibina:2020msd} that the Ayon-Beato-Garcia (ABG) NLED model for Bardeen can support stable null bound orbits outside its horizons\footnote{However, it was shown in~\cite{Novello:2000km} that some ABG proposals suffer from singularity on its null geodesic.}. In this work, we instead focus on the original Bardeen approach in one dimension higher. The metric~\eqref{abg_sol} still solves the $5d$ Einstein's equations. Applying the same rescaling as \eqref{orbitr}, we have the orbit equation as
\begin{figure}[!ht]
	\centering
	\begin{tabular}{cc}
		\includegraphics[height=12cm]{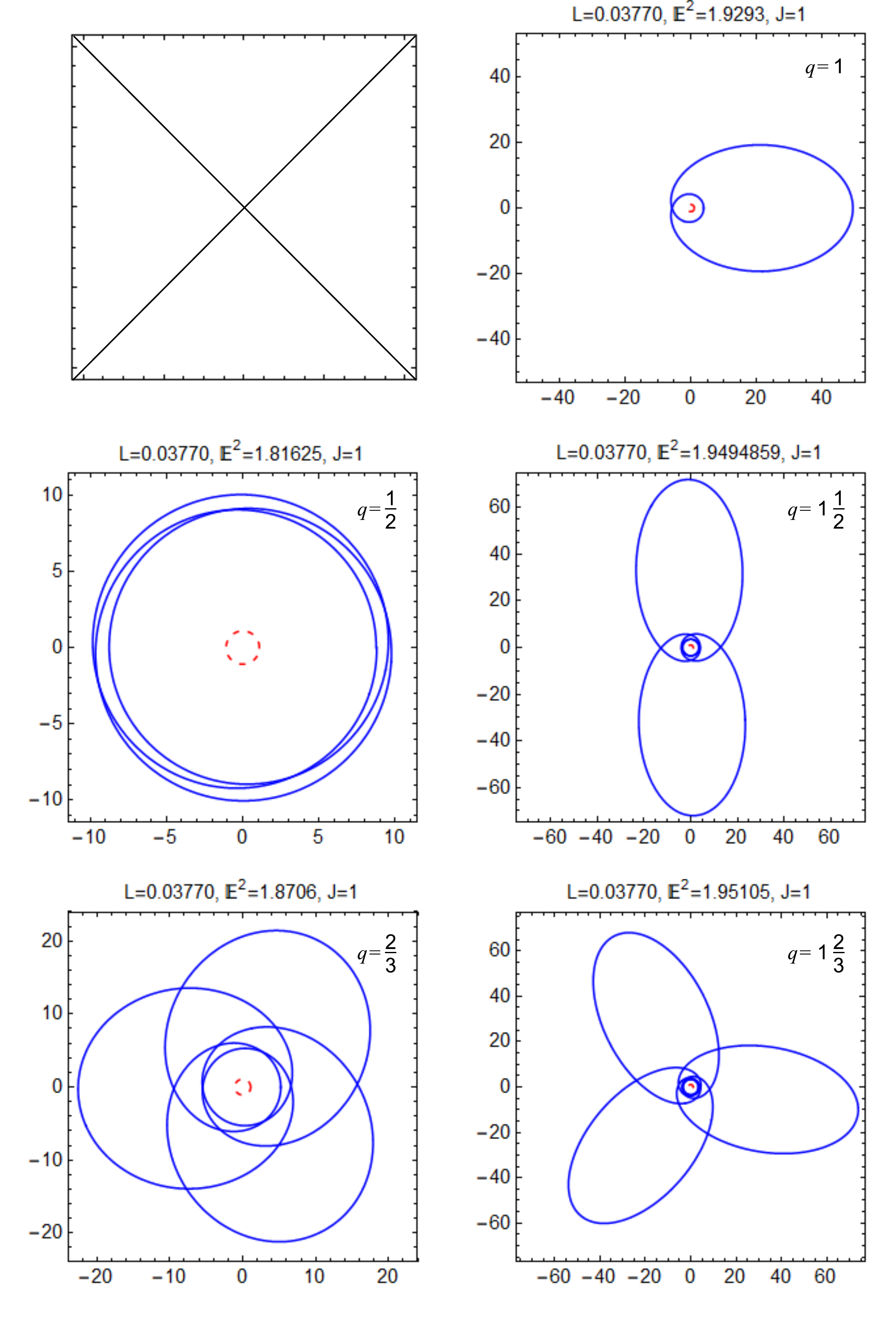}&
		\includegraphics[height=12cm]{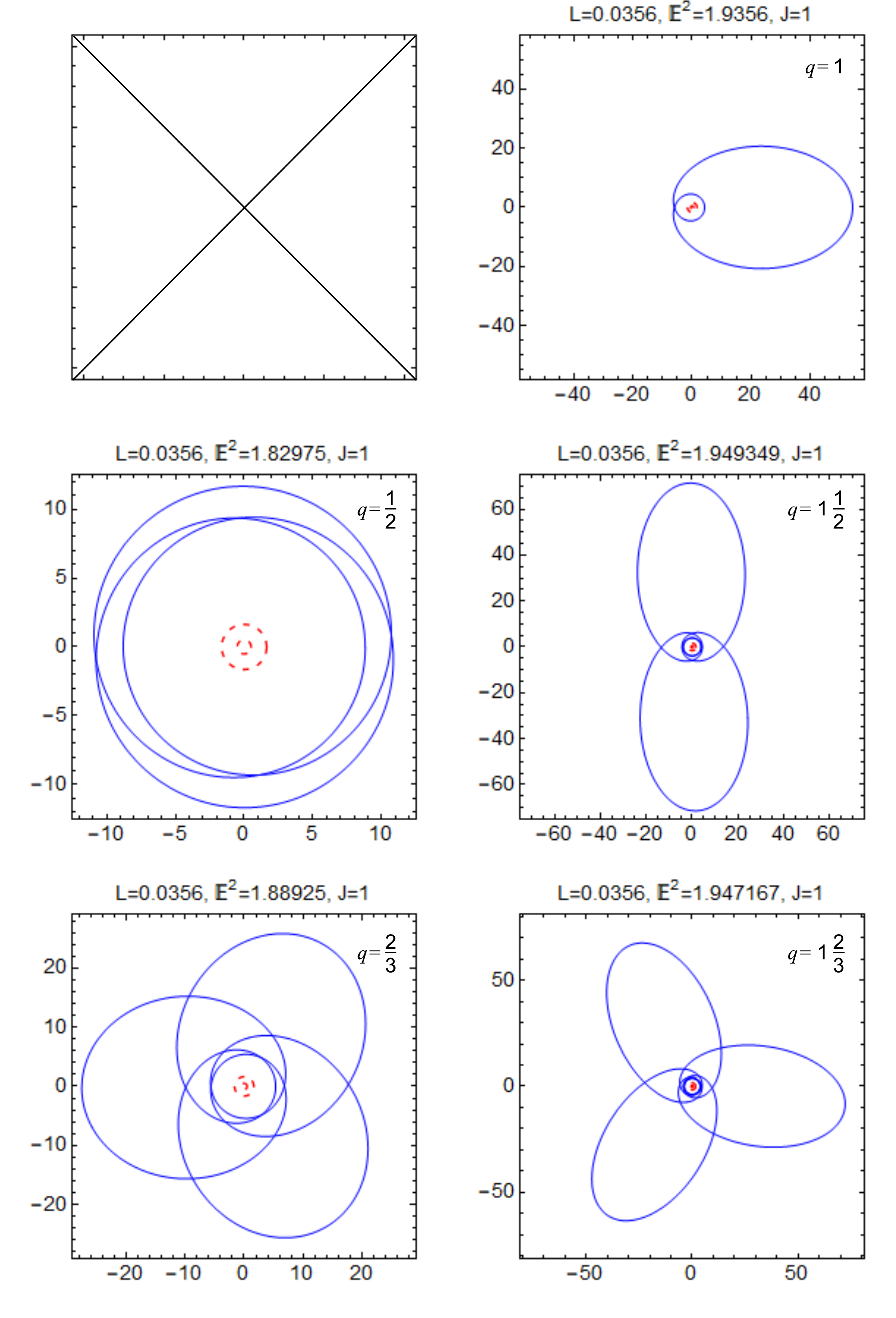}
	\end{tabular}			
	\caption{The $z=1,2,3$ orbits in the timelike condition: extremal case with $w=0$ for the first column and $w=1$ for the second column; two-horizon case with $w=0$ for the third column and $w=1$ for the fourth column. Note that the first entries in the first and third column are blank because the $q=0+\frac{0}{1}$ orbits are inaccessible.}
	\label{b_t}
\end{figure}
Unlike the RN case, we have been unable to cast Eq.~\eqref{geo} into a form that can be solved analytically,
\begin{equation}
\label{borbitr}
\bigg(\frac{dr}{d\phi}\bigg)^2 = \frac{2 \left(r^6 L (J+\epsilon )+r^4\right)}{\left(Q^2+r^2\right)^{3/2}}+r^4 L \left(\mathbb{E}^2+J+\epsilon \right)-r^2.
\end{equation}
To solve it numerically in Mathematica, it is better to rewrite it in terms of the $2^{nd}$-order ODE:
\begin{eqnarray}
\label{borbitr2}
\frac{d^2 r}{d\phi^2} &=& \frac{3 L r^7 (J+\epsilon)}{\left(Q^2+r^2\right)^{5/2}}+\frac{3 r^5 \left(2 Q^2 L (J+\epsilon )-1\right)}{\left(Q^2+r^2\right)^{5/2}} \nonumber \\
&& + r^3 \left(\frac{4}{\left(Q^2+r^2\right)^{3/2}}-2 L \left(-\mathbb{E}^2+J+\epsilon \right)\right) -r.
\end{eqnarray}

\begin{figure}[!ht]
	\centering
	\begin{tabular}{cc}
		\includegraphics[height=12cm]{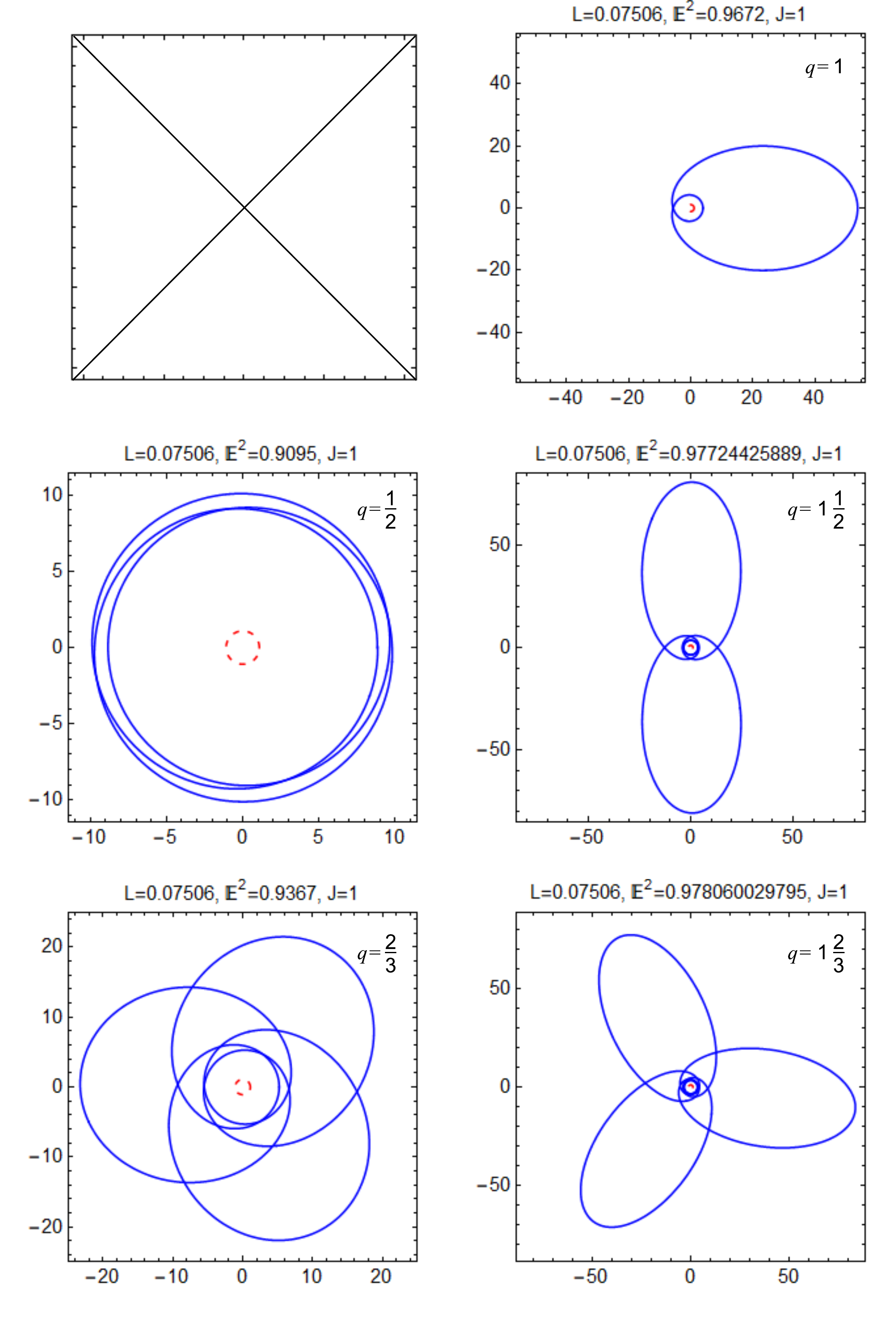}&
		\includegraphics[height=12cm]{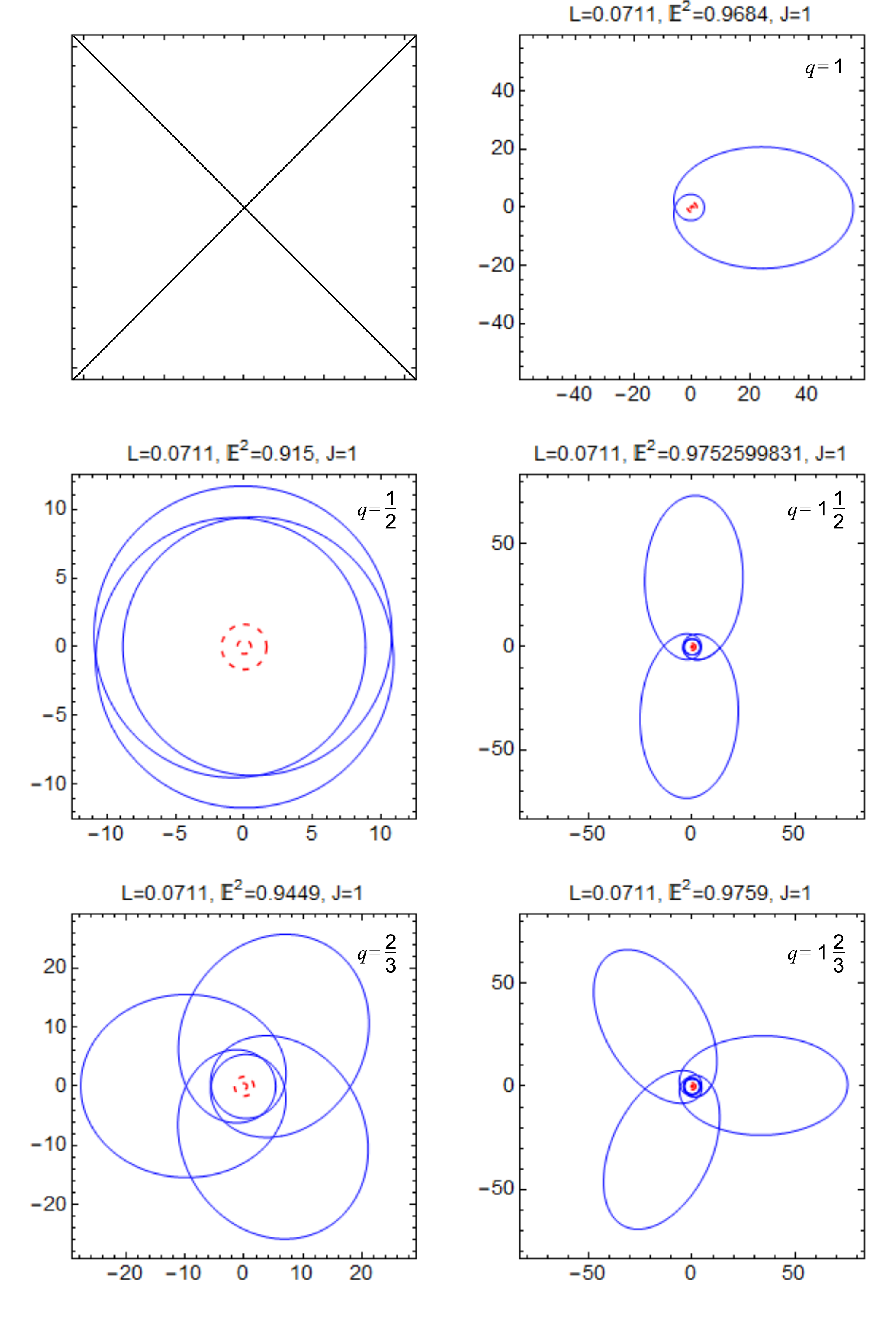}
	\end{tabular}			
	\caption{The $z=1,2,3$ orbits in the null condition: extremal case with $w=0$ for the first column and $w=1$ for the second column; two-horizon case with $w=0$ for the third column and $w=1$ for the fourth column. Note that the first entries in the first and third column are blank because the $q=0+\frac{0}{1}$ orbits are inaccessible.}
	\label{b_n}
\end{figure}
We obtain the numerical solutions of periodic orbits, both for timelike as well as null conditions. They are shown up to $q=1{2\over3}$ in Figs.~\ref{b_t}-\ref{b_n}, respectively. For each case we are able to obtain the extremal and two-horizon solutions. What is interesting is that the orbits found in extremal and two-horizon case have almost identical form for each equivalent counterpart. %Even though the  Bardeen black hole has never been inspected with this taxonomy, it is safe to say that all timelike orbits shown here would also exist in its 4-dimensional form ($J=0$), in the same manner of how the periodic orbits of RN exist in both its black hole and black string versions. %For the null case, however, the situation is different. Our results do not have its $4d$ counterpart, as can be inf

\begin{table}[!h]
	\setlength{\tabcolsep}{1.2em}
	\begin{tabular}{cccc}
		\begin{tabular}{||c |} 
			\hline
			$q$ \\ \hline
			SCO \\ \hline
			1/2 \\ \hline
			2/3 \\ \hline
			1     \\ \hline
			1 1/2 \\ \hline
			1 2/3 \\ \hline
			UCO \\ \hline
		\end{tabular}
		\begin{tabular}{|c | c |} 
			\hline
			$\mathbb{E}$ (Extremal) & Position \\ \hline
			1.815650000 & 0.00\% \\ \hline
			1.816250000 & 0.44\% \\ \hline
			1.870600000 & 39.88\% \\ \hline
			1.929300000 & 82.48\% \\ \hline
			1.949485900 & 97.13\% \\ \hline
			1.951050000 & 98.27\% \\ \hline
			1.953438690 & 100.00\% \\ \hline
		\end{tabular}
		\begin{tabular}{|c | c |} 
			\hline
			$\mathbb{E}$ (Two-Horizon) & Position \\ \hline
			1.827332000 & 0.00\% \\ \hline
			1.829750000 & 1.95\% \\ \hline
			1.889250000 & 49.86\% \\ \hline
			1.935600000 & 87.18\% \\ \hline
			1.949349000 & 98.25\% \\ \hline
			1.950209151 & 98.94\% \\ \hline
			1.951519700 & 100.00\% \\ \hline
		\end{tabular}
		\begin{tabular}{|c ||} 
			\hline
			Deviation \\ \hline
			- \\ \hline
			1.07\% \\ \hline
			7.06\% \\ \hline
			3.32\% \\ \hline
			0.79\% \\ \hline
			0.48\% \\ \hline
			- \\ \hline
		\end{tabular}
	\end{tabular}
	\caption{Comparison of energy level and the position of particular orbit inside the energy range as percentage for timelike case in extremal and two-horizon nonsingular black string. SCO refers to the bottom level of energy range (no whirl), while UCO refers to the top one (maximum whirl).}
	\label{bt12}
\end{table}
To create a more comprehensive view of periodic orbits in the nonsingular string, we provide similar tables as in \ref{rn_taxo}. Tables \ref{bt12} and \ref{bn12} show various energy levels and each of their corresponding position in extremal and two-horizon cases, respectively. A quick calculation of average of standard deviation of each case would give $0.09\%$ for extremal and $0.07\%$ for two horizon one. This is way smaller than the ones found in RN black string, which means the prediction for the position of each periodic orbit will be more precise in the nonsingular black string. 
\begin{table}[!hb]
	\setlength{\tabcolsep}{1.2em}
	\begin{tabular}{cccc}
		\begin{tabular}{||c |} 
			\hline
			$q$ \\ \hline
			SCO \\ \hline
			1/2 \\ \hline
			2/3 \\ \hline
			1     \\ \hline
			1 1/2 \\ \hline
			1 2/3 \\ \hline
			UCO \\ \hline
		\end{tabular}
		\begin{tabular}{|c | c |} 
			\hline
			$\mathbb{E}$ (Extremal) & Position \\ \hline
			0.908371200 & 0.00\% \\ \hline
			0.908659500 & 0.41\% \\ \hline
			0.936700000 & 39.96\% \\ \hline
			0.967200000 & 82.98\% \\ \hline
			0.977244258 & 97.14\% \\ \hline
			0.978060029 & 98.29\% \\ \hline
			0.979268855 & 100.00\% \\ \hline
		\end{tabular}
		\begin{tabular}{|c | c |} 
			\hline
			$\mathbb{E}$ (Two-Horizon) & Position \\ \hline
			0.913804300 & 0.00\% \\ \hline
			0.915000000 & 1.91\% \\ \hline
			0.944900000 & 49.68\% \\ \hline
			0.968400000 & 87.23\% \\ \hline
			0.975259983 & 98.19\% \\ \hline
			0.975842810 & 99.12\% \\ \hline
			0.976394418 & 100.00\% \\ \hline
		\end{tabular}
		\begin{tabular}{|c ||} 
			\hline
			Deviation \\ \hline
			- \\ \hline
			1.06\% \\ \hline
			6.88\% \\ \hline
			3.01\% \\ \hline
			0.74\% \\ \hline
			0.58\% \\ \hline
			- \\ \hline
		\end{tabular}
	\end{tabular}
	\caption{Comparison of energy level and the position of particular orbit inside the energy range as percentage for null case in extremal and two-horizon Bardeen black string. SCO refers to the bottom level of energy range (no whirl), while UCO refers to the top one (maximum whirl).}
	\label{bn12}
\end{table}
Another point to be noted is that the average of deviation for timelike and null condition is $2.54\%$ and $2.45\%$, respectively. This is a similar manner found in RN black string: the distance between energy levels is more consistent when we compare each specific horizon case, which means this behavior exists generally on charged black holes. By carefully inspecting the numbers from both table, we also find that the orbits in two-horizon condition are more dense toward the local maximum (point of UCO). This behavior is observed in both timelike and null orbits. Nevertheless, one might notice that in the two-horizon nonsingular black string the shift is less drastic compared to similar behavior in two-horizon RN \ref{rn_taxo} where the orbits become highly condensed near the maximum.

As in the timelike RN, the accumulated radial angle can be calculated as follows:
\begin{equation}
\Delta\varphi_r= 2\int_{r_p}^{r_a}\frac{\dot{\phi}}{\dot{r}}dr = 2\int_{r_p}^{r_a} \frac{dr}{Lr^2\sqrt{\mathbb{E}^2-\left(1-\frac{2 M r^2}{(r^2 + Q^2)^{3/2}}\right)\left(\epsilon +J + \frac{1}{L r^2} \right)}}.
\end{equation}
We calculate numerically the accumulated angle and show it in Table~\ref{baracc}. We compare the angles we calculate with the corresponding angle from the $4d$ counterpart. As in the RN case, the nonsingular black strings exhibits higher values for the accumulated angle. From the Table \ref{baracc} the $4d$ angles reach maximum at $q=1{1\over2}$, then decreases. the $5d$ counterpart, however, keeps increasing as $q$ increasing.
\begin{table}[!ht]
	\setlength{\tabcolsep}{1.5em}
	\centering
	\begin{tabular}{||c||c|c||c|c||}
		\hline
		q &   $\Delta\varphi_{r-2\ horizon}$ & $\Delta\varphi_{r-extremal}$ & $\Delta\varphi_{r-4d extremal}$ \\ \hline
		1/2 &  50.3815 & 49.8285  & - \\ \hline
		2/3 &  55.5692 & 54.0185  & 37.9724 \\ \hline
		1 &  66.6127 & 64.715  & 44.4775 \\ \hline
		1 1/2 &  83.3911 & 81.0931 & 50.7586 \\ \hline
		1 2/3 &  87.6954 & 85.7237 &  48.5055 \\ \hline
	\end{tabular}
	\caption{Accumulated angles for the nonsingular black strings. The second and third columns are the angles for 2-horizon and extremal string, respectively. The fourth column is the corresponding angles in the $4d$ extremal Bardeen BH from~\cite{Gao:2020wjz}. The angle for $q=1/2$ in $4d$ extremal Bardeen is empty because the authors do not show solutions for $(2,0,1)$.}
\label{baracc}
\end{table}

\section{Conclusion}
\label{con}

This works deals with the bound orbits of massive and massless particles around charged RN and nonsingular black strings. These toy models are quite simple, but the geodesics are rich enough to have genuine observables that can be distinguished from their $4d$ counterparts. Following Hackmann~\cite{Hackmann:2010tqa} the RN bound orbits can be expressed analytically in terms of the Weierstrass function, while the nonsingular string must be solved numerically. We present solutions for timelike and null bound orbits in both models, characterized by the rational number $q$. We investigate bond orbit solutions for several $q$ up to $q=1{2\over3}$. Higher values of $q$ can, in principle, be obtained easily.

What is novel here are twofold. First, the existence of stable bound orbits for photon which does not exist in its 4d counterpart. The addition of (trivial) extra dimension modifies the geodesic equations and thus the shapes of the corresponding $V_{eff}$. As can be seen from Figs.~\ref{rn_vn} and \ref{b_vn}, the non-zero $J$ raises the tail of $V_{eff}$ to create a local maxima outside the corresponding horizons. This is confirmed, for example in the case of RN, by Eq.~\eqref{nullco} where the existence of $j\equiv JL$ gives a cubic term in the circular orbit condition. This another extremum turns out to be minimum and lies outside the corresponding horizon(s). Thus this extra bound orbit is stable and observable. For the extremal RN the roots for the cubic equation can be found analytically, as shown in Eq.~\eqref{exco}. All $r_{CO}$s are located outside or on the extremal horizon, provided $0<j<1/8M^2$.

For the massive case, we show that the circular orbits require higher particle's energy and angular momentum (remember that the angular momentum $\mathbb{L}\equiv1/\sqrt{L}$). The ISCO radii are indistinguishable from its $4d$ counterpart. However, as pointed out by authors of~\cite{Babar:2017gsg}, the accumulated angle of the timelike orbit might provide observational signature that can distinguish these black strings from their corresponding black hole counterparts. In Tables~\ref{accRNext}-\ref{accRNnext} and~\ref{baracc} we show comparison of radial accumulated angles for black strings and black holes, both for the RN and the nonsingular, respectively. In all cases, the black string accumulated angles are higher compared to the black holes. Thus, measuring such angle for bound orbits of stars around some black hole might give a hint on whether our universe is higher-dimensional or not.

It is probably worthwhile to comment on the size of extra dimension. While the existence of black string solution itself is oblivion to the extra-dimensional size, stability prefers it to be compact, not infinite~\cite{Gregory:1987nb, Harmark:2007md}. The bounds comes from particle physics as well as cosmology. From particle physics, recent report by CMS collaboration suggests that the size for one ``large" extra dimension based on the Randall-Sundrum scenario~\cite{Randall:1999ee, Randall:1999vf} is of order $\sim10^{-4}$ fm~\cite{CMS:2022pjv}. A much more optimistic bounds comes from cosmology. Upper bounds value for the emission ring and the angular shadow diameter from the EHT observation on SgrA* indicates that the extra dimension might be as large as $\sim0.4$ mm~\cite{Tang:2022hsu}.

In this work we only consider static charged black strings in the simplest extension. It would be interesting to extend our investigation to black strings endowed with NLED charged, to regular black strings beyond the Bardeen solutions, or to black strings possessing Lorentz boost as in~\cite{Gregory:1995qh}. We are addressing those issues in the forthcoming works.

\acknowledgments

We thank Heribertus Hartanto and Byon Jayawiguna for the enlightening discussions, Adolfo Cisterna for informing us about Refs.~\cite{Cisterna:2020rkc, Cisterna:2021ckn}, Reinard Primulando for the Ref.~\cite{CMS:2022pjv}, and Jason Kristiano for helping us with the Ref.~\cite{Gao:2020wjz}. This work is supported by grants from Universitas Indonesia through Hibah Riset PPI Q1 No. NKB-583/UN2. RST/HKP.05.00/2021.

\section*{Data Availability Statement}

Data sharing is not applicable to this article as no data sets were generated or analyzed during the current study.

%===========================================================

\end{document}